\documentclass[pdftex,twocolumn,epjc3]{svjour3}          % twocolumn

\RequirePackage[T1]{fontenc}

\smartqed  % flush right qed marks, e.g. at end of proof

\RequirePackage{graphicx}
\RequirePackage{mathptmx}      % use Times fonts if available on your TeX system
\RequirePackage{flushend}
\RequirePackage[numbers,sort&compress]{natbib}
\RequirePackage[colorlinks,citecolor=blue,urlcolor=blue,linkcolor=blue]{hyperref}

\usepackage{amsmath}
\usepackage{graphicx}
\usepackage{subfigure}
\journalname{Eur. Phys. J. C}

\begin{document}

\title{Qualifying ringdown and shadow of black holes under general parametrized metrics with photon orbits}

\author{Song Li\thanksref{addr1,addr2,e1}
        \and
        Ahmadjon A. Abdujabbarov\thanksref{addr1,addr3,e2}
        \and
        Wen-Biao Han\thanksref{addr1,addr4,addr2,e3}
}

\thankstext{e1}{lisong20@mails.ucas.ac.cn}
\thankstext{e2}{ahmadjon@astrin.uz}
\thankstext{e3}{Corresponding author: wbhan@shao.ac.cn}

\institute{Shanghai Astronomical Observatory, Shanghai, 200030, China \label{addr1}
          \and
          School of Astronomy and Space Science, University of Chinese Academy of Sciences,
Beijing, 100049, China\label{addr2}
          \and
          Ulugh Beg Astronomical Institute, Astronomicheskaya 33,
	Tashkent 100052, Uzbekistan \label{addr3}
	          \and
	          School of Fundamental Physics and Mathematical Sciences, Hangzhou Institute for Advanced
	          Study, UCAS, Hangzhou 310024, China \label{addr4}
}

\date{Received: date / Accepted: date}

\maketitle

\begin{abstract}
The motion of photons around black holes determines the shape of shadow and match the ringdown properties of a perturbed black hole. Observations of shadows and ringdown waveforms will reveal the nature of black holes. In this paper, we study the motion of photons in a general parametrized metric beyond the Kerr hypothesis. We investigated the radius and frequency of the photon circular orbits on the equatorial plane and obtained  fitted formula with varied parameters. The Lyapunov exponent which connects to the decay rate of the ringdown amplitude is also  calculated. We also analyzed the shape of shadow with full parameters of the generally axisymmetric metric. Our results imply the potential constraint on black hole parameters by combining the Event Horizon Telescope and gravitational wave observations in the future.
\end{abstract}

\section{Introduction}
One of the main problems of gravity theories is to test the theory in strong field regime with high accuracy. However,  the LIGO-Virgo experiments on gravitational wave observation ~\cite{L_V_e1,L_V_e2,L_V_e3} and observation of black hole shadow image at the center of M87 elliptic  galaxy~\cite{M87_Galaxy} give us the opportunity to develop new tests of general relativity and modified/alternative theories of gravity in the strong field regime. Further improvements of the experiments and observations will give more precise results and opportunity to obtain constraints on different theories of gravity. 

Up to now the general relativity proposed by Albert Einstein in 1916 is considered as the main theory of gravity which is justified by the different tests in the weak (e.g. Solar system tests~\cite{S_S1,S_S2}) and strong field (e.g. gravitational waves~\cite{G_W1,G_W2}, shadow of M87~\cite{M87_tests1,M87_tests2,M87_tests3,Khodadi_2020,Psaltis2021,Kocherlakota2021} regimes with high accuracy. However, general relativity meets number of problems connected with the spacetime singularity at the origin of the solutions of the field equations, renormalization, compatibility with quantum field theory and etc. Despite the attempts to resolve the singularity problem introducing conformal transformation~\cite{Conformal_1,Conformal_3,Conformal_4,Conformal_5,Cele_coor_1}, coupling with nonlinear electrodynamics~\cite{Nonlinear_1,Nonlinear_2,Nonlinear_3}, higher dimensional corrections~\cite{Higer_1} and etc,  no single unique theory has been found to resolve all fundamental problems of general theory of relativity. Thus, we need to deal with the big number of the modified/alternative theories of gravity to resolve the fundamental problems of general relativity and describe the current observational and experimental data. 

There is a belief that astrophysical black holes are described by the Kerr spacetime (at least by Schwarzschild spacetime when the effect of rotation is negligible). However, the so-called Kerr hypothesis has not been yet well confirmed by current experimental and observational data. The new parameters of solutions within modified or alternative theories of gravity representing the deflections from Kerr spacetime may mimic the effects of the spin parameter of the Kerr black holes~\cite{Kerr_1,Kerr_2,Kerr_3}. Due to this fact one needs to develop more independent tests of the gravity theory~\cite{G_t_1,G_t_2} together with further improvements of the astrophysical instrumentation~\cite{Instrument_1,Instrument_2}. 

On the other hand big number of alternative and modified theories of gravity and corresponding solutions of field equation describing the compact gravitating object creates the difficulties associated with resolving the cross mimicking of the parameters of theories. One of the attempt to resolve this issue is to use the parametrization of the spacetime metric. The parametrization of the spacetime around rotating black hole may help to cover big number of the solutions of different gravity models. Several ways of parametrization of the spacetime describing the rotating black hole have been proposed by different authors~\cite{Johannsen,KRZ,Ni_2016}. So-called KRZ metric was propsed by Konoplya, Rezzolla, and Zhidenko~\cite{KRZ} where they have suggested parametrization of the spacetime of rotating black hole. They have proposed a parametric frame to describe the spacetime of the axisymmetric black holes which have the Killing vector $\eta = (0,0,0,1)$. The most interesting point of the KRZ metric is that it contains many significant parameters. These parameters associated with physical properties of the black hole and we briefly discuss them in the text of the paper. Particularly, when all the parameters of KRZ parametrization are equal to zero, the specetime metric reduces to Kerr one. 

The most distinguishable feature of the metric theories of gravity is the light deflection and other effects connected with photon motion in the curved spacetime. Some authors studied the photon surfaces\cite{Claudel_2000} through the photon motion. The gravitational lensing was the one of the first consequences of the general relativity and has been discovered by Einstein. The study of the light deflection can be used to study either gravitational object or distant source. The review of the gravitational lensing effect can be found in Refs~\cite{Synge60,Schneider92,Perlick00,Perlick04}. The effects of the plasma on gravitational lensing in different spacetimes have been studied in\cite{Rogers15,Rogers17,Er18,Broderick03,Bicak75, Kichenassamy85,Perlick17,Perlick15,Abdujabbarov17,Eiroa12, Kogan10,Tsupko10,Tsupko12,Morozova13,Tsupko14,Kogan17, Hakimov16,Turimov18,Benavides16,Kraniotis14}. The most basic gravitational lensing is the Schwarzschild lensing and it have been studied in\cite{Virbhadra_1999,Virbhadra_2008}, some papers also studied the cosmic censorship hypothesis (CCH) with the gravitational lensing\cite{Virbhadra_2002,Virbhadra_2007}. 

Additionally, the properties of circular orbits of photons around black holes reflect on the ringdown signal from the merger of binary black holes \cite{Goebel1972ApJ,Ferrari_84}. The latter are quasinormal modes (QNMs) which describe the end state of a black hole-black hole merger. Therefore the gravitational-wave emission at late times can be well described by the properties of null geodesics on unstable circular orbits at the black hole’s light ring \cite{Yang_12,LE}. The direct calculation of QNMs of the KRZ metric is difficult, however, one may use the alternative method to calculate the frequency and Lyapunov exponent of unstable circular orbits leading to the features of ringdown of waveforms \cite{Goebel1972ApJ,Ferrari_84,Yang_12,LE}. Some authors work on the spherical and static parametrized RZ metric and proposed the higher order WKB method to reduce the difficult of computing the QNMs~\cite{RZ_2020}. On the other hand the standard metric perturbations of the Schwarzschild black hole has been studied in the pioneer work of  Regge and Wheeler~\cite{Regge57} and Zerilli~\cite{Zerilli70}. Later the QNM of the black holes have been calculated using the perturbative method in different works (see, e.g.~\cite{Vishveshwara70,Nagar05,Chandrasekhar98,Kokkotas99,Berti09a,Konoplya11} and reference therein). 

The recent~\cite{EHT_1,EHT_2,EHT_3,EHT_4} and future observation of the black hole shadow by Event Horizon Telescope (EHT) using very long baseline interferometry (VLBI) technique can be used to explore the the gravity in the strong field regime around supermassive black hole (SMBH). At the same time one may test the gravity theories using the observational data from the black hole shadow. The shadow of SMBH has been theoretically studied in Refs.\cite{Takahashi05, Hioki09, Amarilla10, Amarilla12, Amarilla13, Abdujabbarov13c, Atamurotov13, Wei13, Atamurotov13b, Bambi15, Ghasemi-Nodehi15, Cunha15, Abdujabbarov15, Atamurotov15a, Ohgami15, Grenzebach15, Mureika17, Abdujabbarov17b, Abdujabbarov16a, Abdujabbarov16b, Mizuno18, Shaikh18b, Kogan17, Perlick17,Schee15, Schee09a, Stuchlik14, Schee09, Stuchlik10} within the different gravity models. Here we plan to study shadow of the black holes described by the parametric spacetime metric proposed in~\cite{KRZ}. 

The paper is organized as follows:
Sect.~\ref{photmotion} is devoted to briefly review of the motion of massive and massless particles in the KRZ space-time and construction of the ray tracing algorithm necessary to investigate the shadow. We also study the frequency of photon orbits~(Sect.~\ref{Frequency}) and the Lyapunov Exponent~(Sect.~\ref{LE}) of light ring. Sect.~\ref{rtcs} describes the ray-tracing code used to construct the shadow of the KRZ metric. In Sect.~\ref{shadow}, we consider the shadow cast by the KRZ space-time for observer at infinity. Finally, in Sect.~\ref{Summary} we summarize the obtained results. Throughout the paper we use a space-like
signature $(-,+,+,+)$, a system of units
in which $G = c = 1$. Greek indices run from $0$ to $3$, Latin indices from $1$ to $3$.

\section{Photon motion \label{photmotion}}
In this section we explore the parametrized KRZ metric proposed in~\cite{KRZ} and investigate the photon motion around compact object described by KRZ metric. The lowest-order metric expression of the KRZ parametrization has the following form:

\begin{eqnarray}\label{metric}
d s^{2}&=&-\frac{N^{2}(\tilde{r}, \theta)-W^{2}(\tilde{r}, \theta) \sin ^{2} \theta}{K^{2}(\tilde{r}, \theta)} d t^{2}\nonumber\\&&-2 W(\tilde{r}, \theta) \tilde{r} \sin ^{2} \theta d t d \phi  \nonumber\\&&+K^{2}(\tilde{r}, \theta) \tilde{r}^{2} \sin ^{2} \theta d \phi^{2}\nonumber\\&&+\Sigma(\tilde{r}, \theta)\left(\frac{B^{2}(\tilde{r}, \theta)}{N^{2}(\tilde{r}, \theta)} d \tilde{r}^{2}+\tilde{r}^{2}d \theta^{2}\right), 
\end{eqnarray}

where $\tilde{r}=r/M, \ \tilde{a}=a / M$ and the other metric functions are defined as~\cite{xin2019}:

\begin{eqnarray}
\Sigma&=&1+a^{2} \cos ^{2} \theta/\tilde{r}^{2}\ ,\\
{N}^{2}&=&\left( 1-{{r}_{0}}/\tilde{r} \right)\nonumber\\&&\left[ 1-{{\epsilon }_{0}}{{r}_{0}}/\tilde{r}+\left( {{k}_{00}}-{{\epsilon }_{0}} \right)r_{0}^{2}/{{{\tilde{r}}}^{2}}+{{\delta }_{1}}r_{0}^{3}/{{{\tilde{r}}}^{3}} \right]\nonumber\\&&+[ {{a}_{20}}r_{0}^{3}/{{{\tilde{r}}}^{3}}+{{a}_{21}}r_{0}^{4}/{{{\tilde{r}}}^{4}}+{{k}_{21}}r_{0}^{3}/{{{\tilde{r}}}^{3}}L]{{\cos }^{2}}\theta\ ,\\
B&=&1+\delta_{4} r_{0}^{2} / \tilde{r}^{2}+\delta_{5} r_{0}^{2} \cos ^{2} \theta / \tilde{r}^{2}\ ,\\
W&=&\left[w_{00} r_{0}^{2} / \tilde{r}^{2}+\delta_{2} r_{0}^{3} / \tilde{r}^{3}+\delta_{3} r_{0}^{3} / \tilde{r}^{3} \cos ^{2} \theta\right] / \Sigma\ ,\\
K^{2}&=&1+a W / r\nonumber\\&&+\left\{k_{00} r_{0}^{2} / \tilde{r}^{2}+k_{21} r_{0}^{3} / \tilde{r}^{3}L \cos ^{2} \theta\right\} / \Sigma\ ,
\\ \label{L}
L&=&\left[1+\frac{k_{22}\left(1-r_{0} / \tilde{r}\right)}{1+k_{23}\left(1-r_{0} / \tilde{r}\right)}\right]^{-1}\ .
\end{eqnarray}

In this paper we use the following parameters defined
as~\cite{xin2019}:

\begin{eqnarray}
r_{0}&=&1+\sqrt{1-\tilde{a}^{2}}, \\  
a_{20}&=&2 \tilde{a}^{2} / r_{0}^{3}, \\ 
a_{21} &=&-\tilde{a}^{4} / r_{0}^{4}+\delta_{6}, \\ \epsilon_{0}&=&\left(2-r_{0}\right) / r_{0}, \\ k_{00}&=&\tilde{a}^{2} / r_{0}^{2} ,\\ 
k_{21}&=&\tilde{a}^{4} / r_{0}^{4}-2 \tilde{a}^{2} / r_{0}^{3}-\delta_{6}, \\ 
w_{00}&=& 2 \tilde{a} / r_{0}^{2}, \\ 
k_{22}&=&-\tilde{a}^{2} / r_{0}^{2}+\delta_{7}, \\ k_{23}&=&\tilde{a}^{2} / r_{0}^{2}+\delta_{8},
\end{eqnarray}
where $r_{0}$  is the radius of the event horizon in the equatorial plane and $\delta_{i}$\  ($i=1,2,3,4,5,6,7,8$) is the dimensionless  parameter describing the corresponding deformation of the parameter in the metric~(\ref{metric}). Particularly, $\delta_{1}$ corresponds to the deformation of $g_{tt}$, $\delta_{2}$ and  $\delta_{3}$ correspond to the deformations of spin, $\delta_{4}$ and $\delta_{5}$ correspond to the deformations of $g_{rr}$, $\delta_{6}$ corresponds to the deformation of the event horizon. In the case when $\delta_{i}=0$ the KRZ one~(\ref{metric}) reduces to Kerr metric and $\tilde{a}=0$ reduces the Kerr metric to 
Schwarzschild one. 

The stationary and axisymmetric KRZ metric is independent of $t$ and $\phi$ coordinates which leads to existence of timelike and spacelike Killing vectors. Consequently, these two Killing vectors correspond to  two conserved quantities: the energy $E$ and the z-component of the angular momentum  $L_z$ of test particle. The conserved energy and angular momentum of the test particle can be expressed as:

\begin{eqnarray}\label{conserved_E}
-E &=g_{t t} \dot{t}+g_{t \phi} \dot{\phi},\\
\Phi &=g_{\phi t} \dot{t}+g_{\phi \phi} \dot{\phi},\label{conserved_L}
\end{eqnarray}

where the overhead dot represents the derivative with respect to the affine parameter (proper time for a massive particle). One can thus express the equation of motion of test particles with these two conserved quantities. Substituting Eqs.~(\ref{conserved_E})-(\ref{conserved_L}) into the normalization condition of the four-velocity $u^\alpha u_\alpha= -1$ for a massive particle, where $u^\alpha = (\dot{t} ,\dot{r},\dot{\theta},\dot{\phi})$ is the 4-velocity, one may obtain the following equation for the motion in the equatorial plane ($\theta=\pi/2$):

\begin{equation}\label{parorb}
g_{tt}\dot{t}^2+g_{rr}\dot{r}^2+g_{\phi\phi}\dot{\phi}^2+2g_{t\phi}\dot{t}\dot{\phi} = -1\ . 
\end{equation}

Similarly one can consider the orbits of photons around black hole. For the photon orbits the normalization condition of the four-velocity take the form $u^\alpha u_\alpha = 0$. Considering the orbit in the equatorial one may get the following expression:

\begin{equation}\label{LRs}
g_{tt}\dot{t}^2+g_{rr}\dot{r}^2+g_{\phi\phi}\dot{\phi}^2+2g_{t\phi}\dot{t}\dot{\phi} = 0\ .
\end{equation}

\begin{enumerate}
    
    \item We expand the normalization equation $u^\alpha u_\alpha = -1$ or $u^\alpha u_\alpha= 0$. 
    
    \item We substitute the equations for the conserved quantities $E$ and  $\Phi$~(\ref{conserved_E})-(\ref{conserved_L}) into the normalization equation. 
    
    \item We rewrite the normalization equation by the two conserved quantities in a form similar to the equation in Newtonian mechanics. 
    
\end{enumerate}

The equations of radial motion containing the effective potential for particle and photon have the following form: 
\begin{eqnarray}\label{Veff}
\frac{E^{2}-1}{2}&=&\frac{1}{2}\dot{r}^{2}+V_{\mathrm{eff}}(r)\ ,
\\
\label{Weff}
\frac{E^{2}}{\Phi^{2}}&=&\frac{1}{\Phi^{2}}\dot{r}^{2}+W_{\mathrm{eff}}(r)\ , 
\end{eqnarray}
where $V_{\mathrm{eff}}$ and $W_{\mathrm{eff}}$ are the effective potential for particle and  photon orbits, respectively. The expressions for the $V_{\mathrm{eff}}$ and $W_{\mathrm{eff}}$ depend on parameters of the chosen spacetime metric. 
However, application of  these steps to the KRZ metric becomes problematic due to its complicate form. Thus here we use a method described in~Ref.~\cite{LRs}. 

\begin{figure*}
\includegraphics[width=0.45 \textwidth]{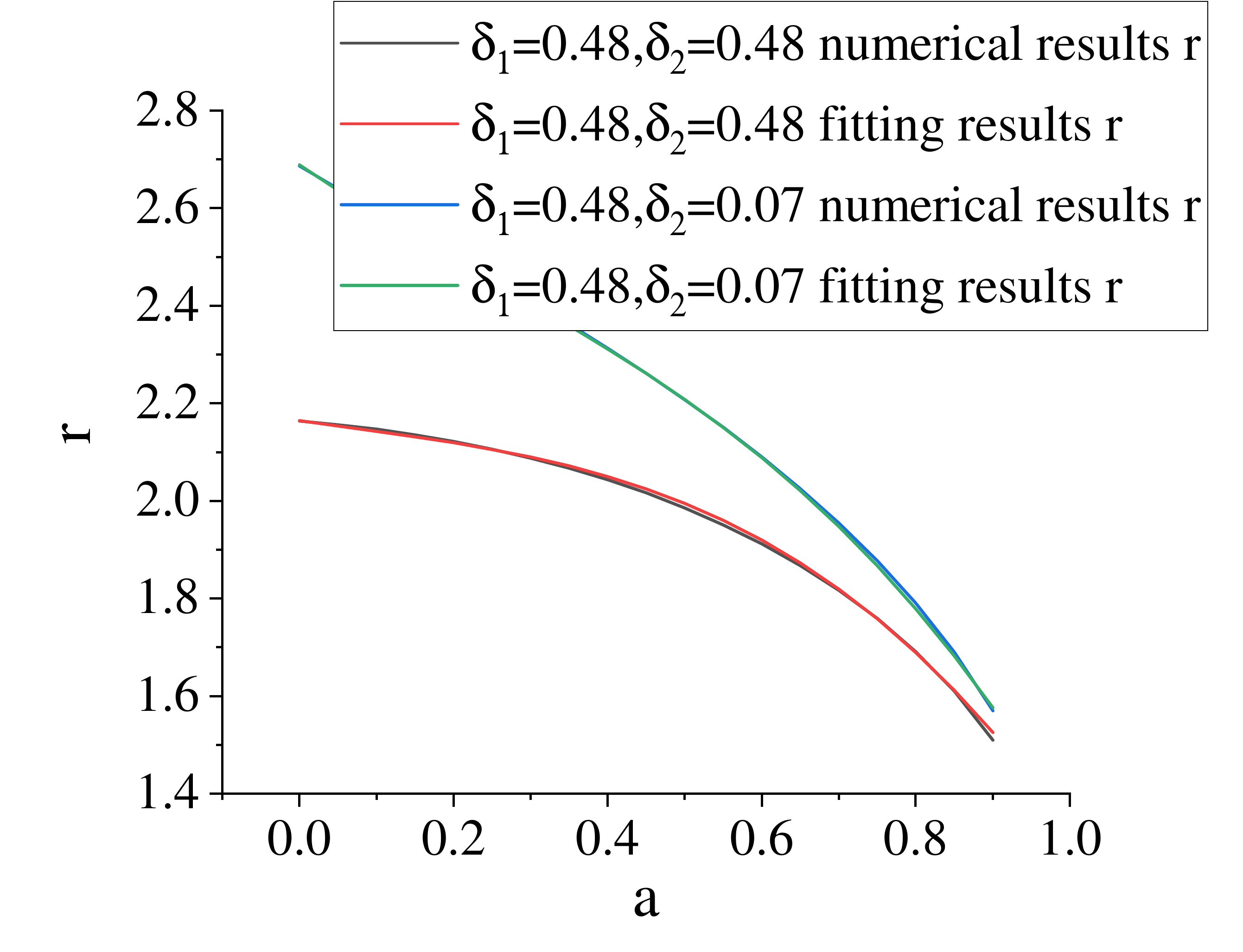}
\includegraphics[width=0.45 \textwidth]{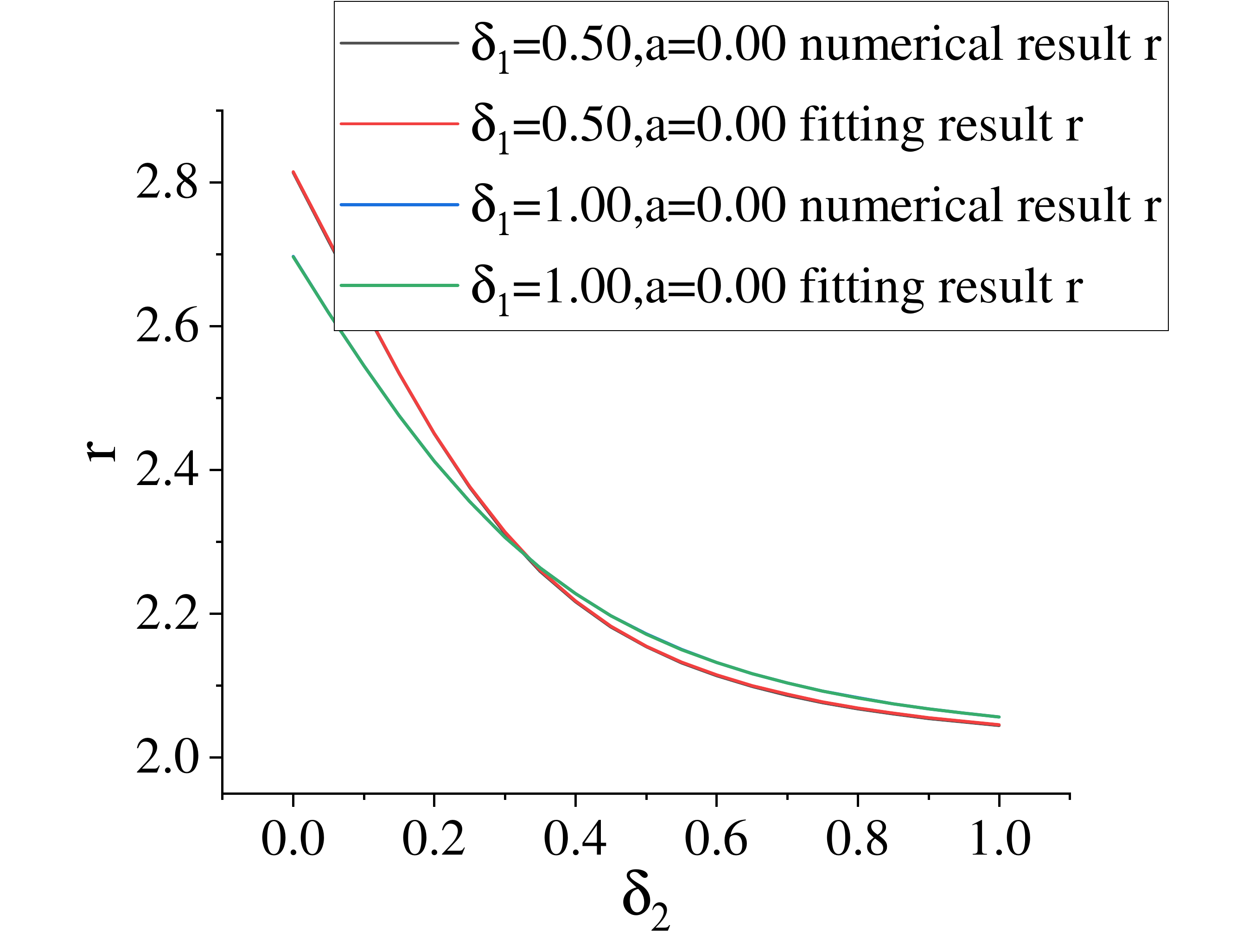}
\caption{Comparison of numerical and fitting results for {the radius} of photon orbit depending on the spin $a$ (left panel) and the $\delta_{2}$ (right panel) parameters for different values of $a$, $\delta_{1}$ and $\delta_{2}$. Note the plane we choose is the equatorial one i.e. $\theta=\pi/2$. \label{fitting_radius}}
\end{figure*}

For the light ring (LR) we have  the following condition 
\begin{equation}
V_{\rm eff}=\nabla V_{\rm eff}=0\ ,
\end{equation}
where the $V_{\rm eff}$ has the following form
\begin{equation}{\label{LR_V}}
V_{\rm eff}=-\frac{1}{D}\left(E^{2} g_{\phi \phi}+2 E \Phi g_{t \phi}+\Phi^{2} g_{t t}\right)\ ,
\end{equation}
One may now easily introduce new potential functions rewriting the Eq.~(\ref{LR_V})in the following form
\begin{equation}
V_{\rm eff}=-\Phi^{2} g_{\phi \phi}\left(\sigma-H_{+}\right)\left(\sigma-H_{-}\right) / D\ ,
\end{equation}
where the $\sigma = 1/b$. $b = |\Phi/E|$ is the inverse impact parameter. 
Newly introduced effective potential $H_{\pm}$ of photon orbits on the orthogonal 2-space has the following form
\begin{equation}
H_{\pm}(r, \theta) \equiv \frac{-g_{t \phi} \pm \sqrt{D}}{g_{\phi \phi}}\ ,
\end{equation}

where $D \equiv\left(g_{t \phi}^{2}-g_{t t} g_{\phi \phi}\right)$. The circular orbits of photon obey the condition  $\partial_{r}H_{\pm}$=0, where the ${\pm}$ sign is associated to the two direction of the rotation. In this paper for our analysis we consider the case corresponding to the  sign "+". Now we can calculate the circular orbits of photon in the equatorial plane i.e.~$\theta=\pi/2$ through the method described in~\cite{LRs}. Since we consider the motion in the equatorial plane and use the condition $\partial_{r}H_{\pm}$=0, one can find out that the circular orbits of photon depend on $\delta_{1}$, $\delta_{2}$ and $\tilde{a}$ only. We cannot get the exact analytical relationship of these three variables $\delta_{1}$, $\delta_{2}$ and $\tilde{a}$, because the relationship of the three variables $\delta_{1}$, $\delta_{2}$ and $\tilde{a}$ in the equation $\partial_{r}H_{\pm}$=0 are complicated. So we try to perform a numerical fit to get the fitting equation $F_{r}(\delta_{1},\delta_{2},\tilde{a})$. After a series of fittings, we get the equation $F_{r}(\delta_{1},\delta_{2},\tilde{a})$ (see \ref{AppA}). Fig.~\ref{fitting_radius} shows the numerical results and fitting results with different variables. From Fig.~\ref{fitting_radius} we can see that the fitting results fits well with the numerical results. With the increase of spin $a$ and $\delta_{2}$, the radius of photon orbits decrease.

After obtaining the radius of the photon circular orbit, one can explore the photon motion. The geodesic equations for null geodesics has the following form:
\begin{equation}\label{Null_equation}
\frac{\mathrm{d}^{2} x^{\mu}}{\mathrm{d} \lambda^{2}}+\Gamma_{\nu \tau}^{\mu} \frac{\mathrm{d} x^{\nu}}{\mathrm{d} \lambda} \frac{\mathrm{d} x^{\tau}}{\mathrm{d} \lambda}=0\ ,
\end{equation}
where $\lambda$ is the affine parameter, $\Gamma_{\nu \tau}^{\mu}$ are Christoffel Symbols defined as 
\begin{equation}\label{Christoffel}
\Gamma_{\nu \tau}^{\mu}=\frac{1}{2} g^{\mu \sigma}\left(g_{\sigma \nu, \tau}+g_{\sigma \tau, \nu}-g_{\nu \tau, \sigma}\right)\ ,
\end{equation}
From Eqs.~(\ref{Null_equation})-(\ref{Christoffel}), one can easily get a differential equations including the terms ${{{d}^{2}}t}/{d{{\lambda }^{2}}}\;,{{{d}^{2}}r}/{d{{\lambda }^{2}}}\;,{{{d}^{2}}\theta }/{d{{\lambda }^{2}}}\;,{{{d}^{2}}\phi }/{d{{\lambda }^{2}}}\;$. Using the relation for radius of photon orbits and the Eqs.~(\ref{conserved_E}), (\ref{conserved_L}) and (\ref{LRs}), one can determine the whole set of initial conditions. 

\begin{figure*}
\centering
\subfigure{
\includegraphics[width=0.45 \textwidth]{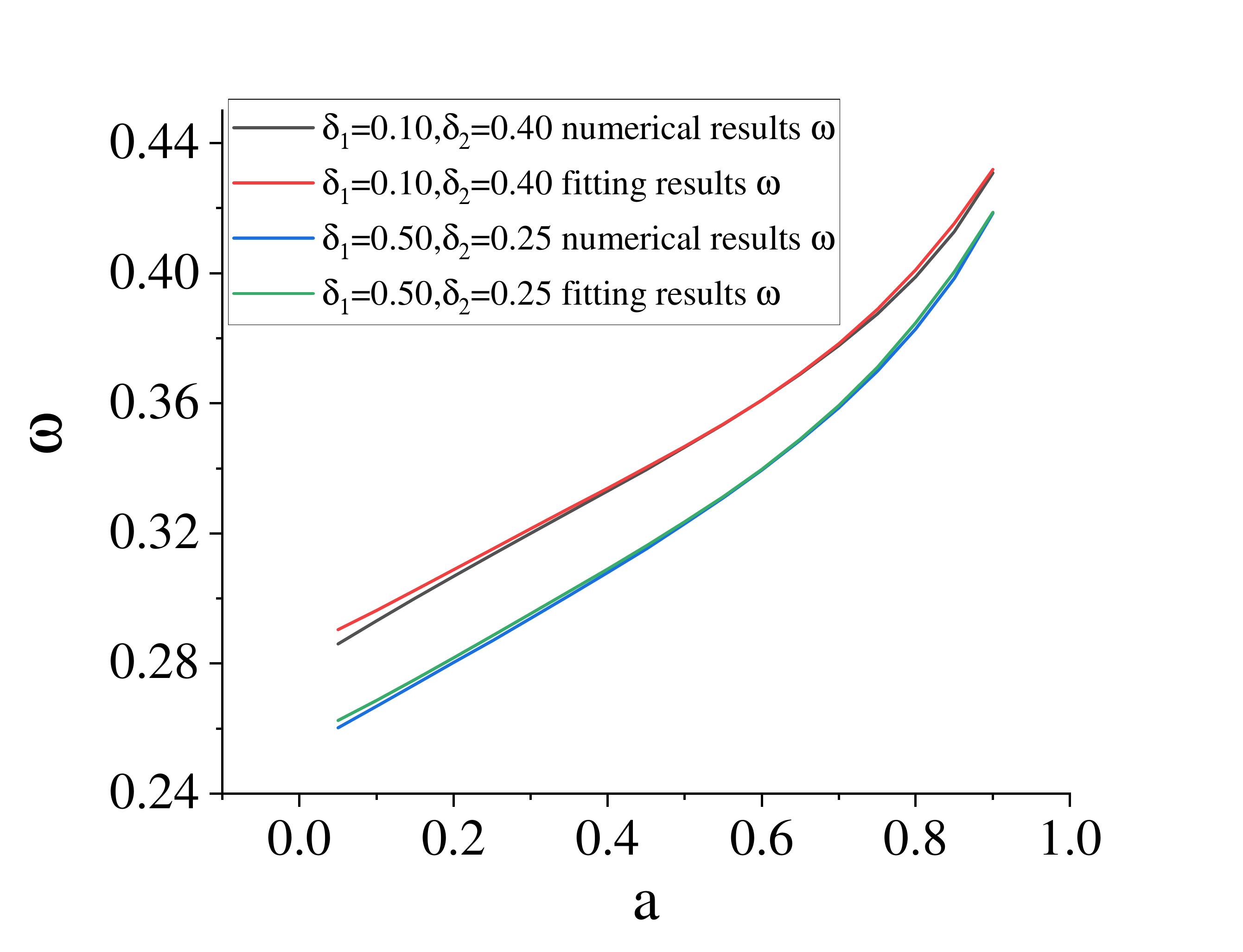}
}
\quad
\subfigure[]{
\includegraphics[width=0.45 \textwidth]{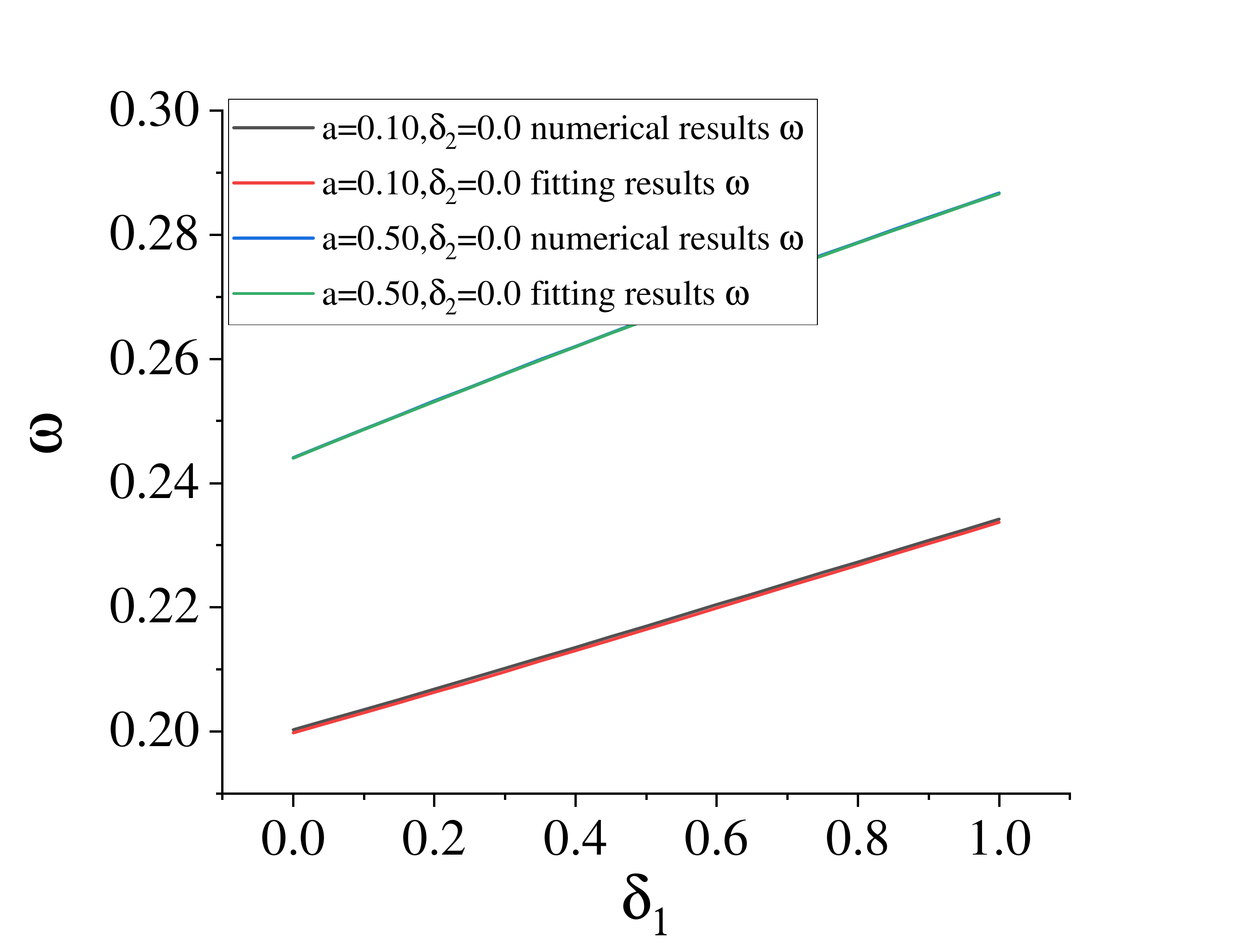}
}
\quad
\subfigure[]{
\includegraphics[width=0.45 \textwidth]{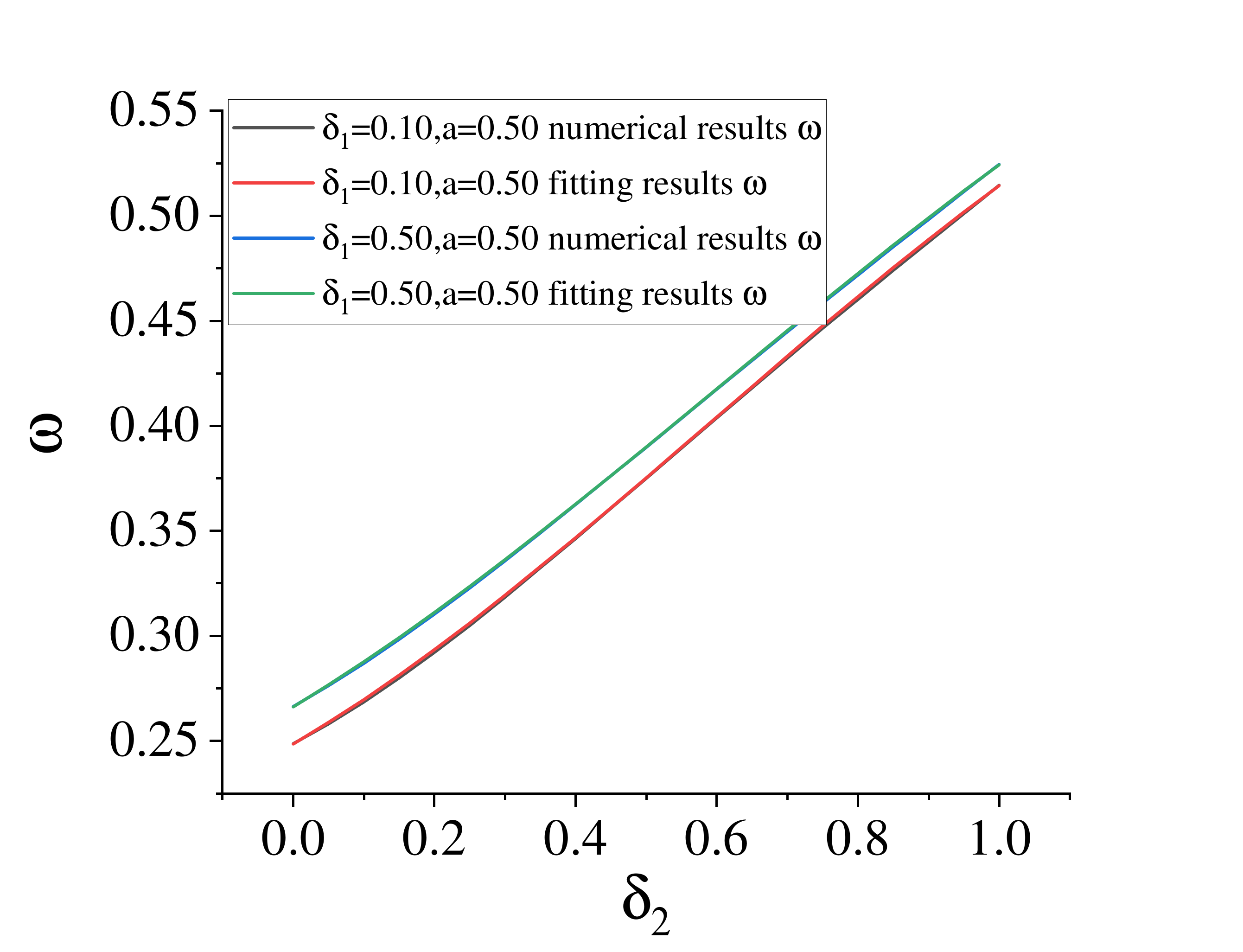}
}
\quad
\subfigure[]{
\includegraphics[width=0.45 \textwidth]{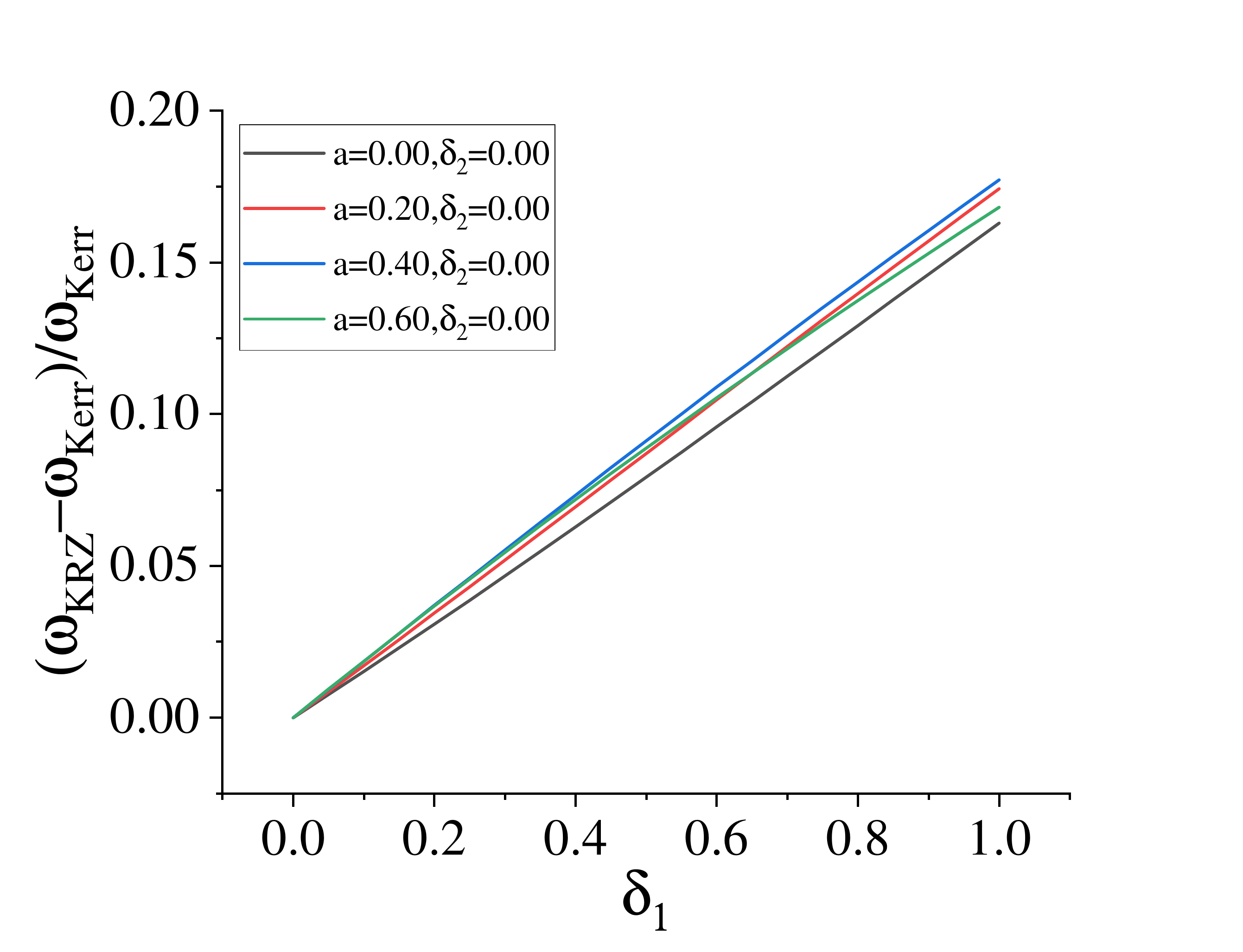}
}
\caption{Comparison of numerical and fitting results for the frequency depending on the spin $a$, $\delta_{1}$ and $\delta_{2}$ parameters. The corresponding parameters have been given in the figures. In all calculations we use the equatorial plane i.e. $\theta=\pi/2$. \label{fitting_frequency}}
\end{figure*}

\section{Frequency of the photon orbits \label{Frequency}}
As we have mentioned before, the QNM of a perturbative black hole can be determined by the unstable circular orbits of photons around the black hole. The frequency of QNM is related the orbital frequency of the light ring \cite{Goebel1972ApJ}. Following the procedure described in Ref.~\cite{LE}, we calculate the frequency of light ring in order to quantify the frequency of ringdown waveforms from KRZ black holes. 

In this subsection, we will explore the frequency of the photon orbits in the equatorial plane around black hole described the KRZ spacetime metric. One can calculate the frequency of photon orbits around Kerr black hole in the following way. Using the normalization condition of the four-velocity $u^\alpha u_\alpha = 0$ one may easily write the radial equation of motion in the form:
\begin{equation}{\label{V_Kerr}}
	\frac{1}{\Phi^{2}}\left(\frac{d r}{d \lambda}\right)^{2}=\frac{1}{b^{2}}-W_{\mathrm{eff}}(r, b, \sigma_m)\ ,
\end{equation}

where $\sigma_m\equiv sign(\Phi)$ and photon effective potential $W_{\mathrm{eff}}(r, b, \sigma_m)$ has the following form:
\begin{equation}
	W_{\mathrm{eff}}(r, b, \sigma_m)=\frac{1}{r^{2}}\left[1-\left(\frac{a}{b}\right)^{2}-\frac{2 M}{r}\left(1-\sigma_m \frac{a}{b}\right)^{2}\right]\ .
\end{equation}
Solving the equation $\partial W_{\mathrm {eff }} /\left.\partial r\right|_{r_{\mathrm {cir}}}=0$ and using the value for $r_{\rm cir}$ one can easily get the the angular velocity $\Omega=\mathrm{d} \phi / \mathrm{dt}$ and consequently get information about the frequency. We may apply the same method to calculate the frequency for the KRZ spacetime.

From the Eqs.~(\ref{conserved_L}) and (\ref{LRs}) one may get the following equation
\begin{equation}{\label{V_KRZ}}
	\frac{1}{\Phi^{2}}\left(\frac{d r}{d \lambda}\right)^{2}=-\frac{g_{t t}+g_{\phi \phi} \Omega^{2}+2 g_{t \phi} \Omega}{g_{r r}\left(g_{t \phi}^{2}+g_{\phi \phi}^{2} \Omega^{2}+2 g_{t \phi} g_{\phi \phi} \Omega\right)}\ .
\end{equation}

Comparing the Eq.~(\ref{V_Kerr}) with Eq.~(\ref{V_KRZ}), one can see that the Eq.~(\ref{V_KRZ}) does not contain the term $1/b^{2}$. This is due to fact that the parameter $b$ is independent of the radial coordinate $r$. Consequently, one can include the parameter $b$ into the expression for $W_{\mathrm {eff}}$ in order to  calculate the frequency.  Now we solve the equation  
$$\left.\frac{\partial W_{\mathrm {eff }}}{\partial r}\right|_{r_{\mathrm {cir}}}=0\ ,$$ 
using the value $r_{\mathrm {cir}}$ obtained using the fitting and get the frequency of the circular photon orbits in KRZ spacetime in the equatorial plane. Since we consider the equatorial plane the frequency of photon orbits depend only on $\delta_{1}$, $\delta_{2}$ and $\tilde{a}$ under the condition $\delta_{4}=0$. It is easy to find the frequency of the photon orbits for the different fixed values of $\delta_{1}$, $\delta_{2}$ and $\tilde{a}$. However, the analytical expression  describing the relation of radius with $\delta_{1}$, $\delta_{2}$ and $\tilde{a}$ cannot be found explicitly due to complicated view of the metric functions. Here we tried to perform a numerical fit to obtain the equation $F_{\omega}(\delta_{1},\delta_{2},\tilde{a})$ for the frequency of photon orbits. After a number of fittings, we have obtained the expressions $F_{\omega}(\delta_{1},\delta_{2},\tilde{a})$ (see \ref{AppB}). Fig.~\ref{fitting_frequency} shows the numerical and fitting results  for the frequency of photon orbits. Although the data shows some differences on the graphs, the relative error between the data is found to be less than one percent after calculation.
From Fig. ~\ref{fitting_frequency} we can find that when $\delta_{1,2}$ is as large as 0.5, the frequency changes less than 10\%. Therefore, only when these non-Kerr parameters are large enough, we can read the derivation from the frequency of ringdown waveform.

\section{The Lyapunov Exponent of Light Ring\label{LE}}

The Lyapunov exponent (LE) is the key indicator of chaos in dynamical systems. Interestingly, the LE of the unstable circular orbit of photon corresponds to the decay rate of the ringdown amplitude from a perturbed black hole \cite{Ferrari_84}. The Lyapunov coefficient characterizes the rate of divergence of nearby null geodesics. Based on the orbital frequency and LE of light-ring, analytical black hole binary merger waveforms are constructed in \cite{LE}, and the waveform amplitude decays as $|\Psi_4| = A_p {\rm sech}[\gamma(t-t_p)]$, where $A_p$ is the amplitude when the congruence converges. In this section we compute the LE of the the photon circular motion around a generally axisymmetric black hole which is described by the KRZ metric.

Authors of Ref.~\cite{Wuxin} have considered two particle's orbital motions with small differences in their initial conditions, then using the metric functions they have calculated LE. However, in this paper we use another approach proposed by McWilliams~\cite{LE}. We consider the perturbation in the radial direction and use the following expression for LR radius~\cite{LE}: 
\begin{equation}\label{LE1}
	r={{r}_{lr}}\left[ 1+\varepsilon \Upsilon \left( t-{{t}_{p}} \right) \right]\ ,
\end{equation}
where $t_{p}$ is the time when the congruence converges, $\varepsilon$ is a small dimensionless order-counting parameter and $r_{lr}$ is the light ring radius. $\Upsilon$ in Eq.~(\ref{LE1}) is a function defined as:
\begin{equation}\label{LE2}
	\Upsilon=\sinh \left[\gamma\left(t-t_{p}\right)\right]\ ,
\end{equation}
where $\gamma$ is the Lyapunov exponent. 
Since we consider the motion in the equatorial plane we consider the effect of the paramaters $\delta_{1}$ and $\delta_{2}$ in the KRZ metric Eq.~(\ref{metric}).

Using the Eqs.~(\ref{LE1})-(\ref{LE2}) one may explore the dependence of LE from the metric parameters. For the appropriate time we plan to plot three-dimensional graphics and explore the influence of the parameters $\delta_{1}$ and $\delta_{2}$ of the metric Eq.~(\ref{metric}) on LE.

\begin{figure*}
\centering
		\includegraphics[width=0.40 \textwidth]{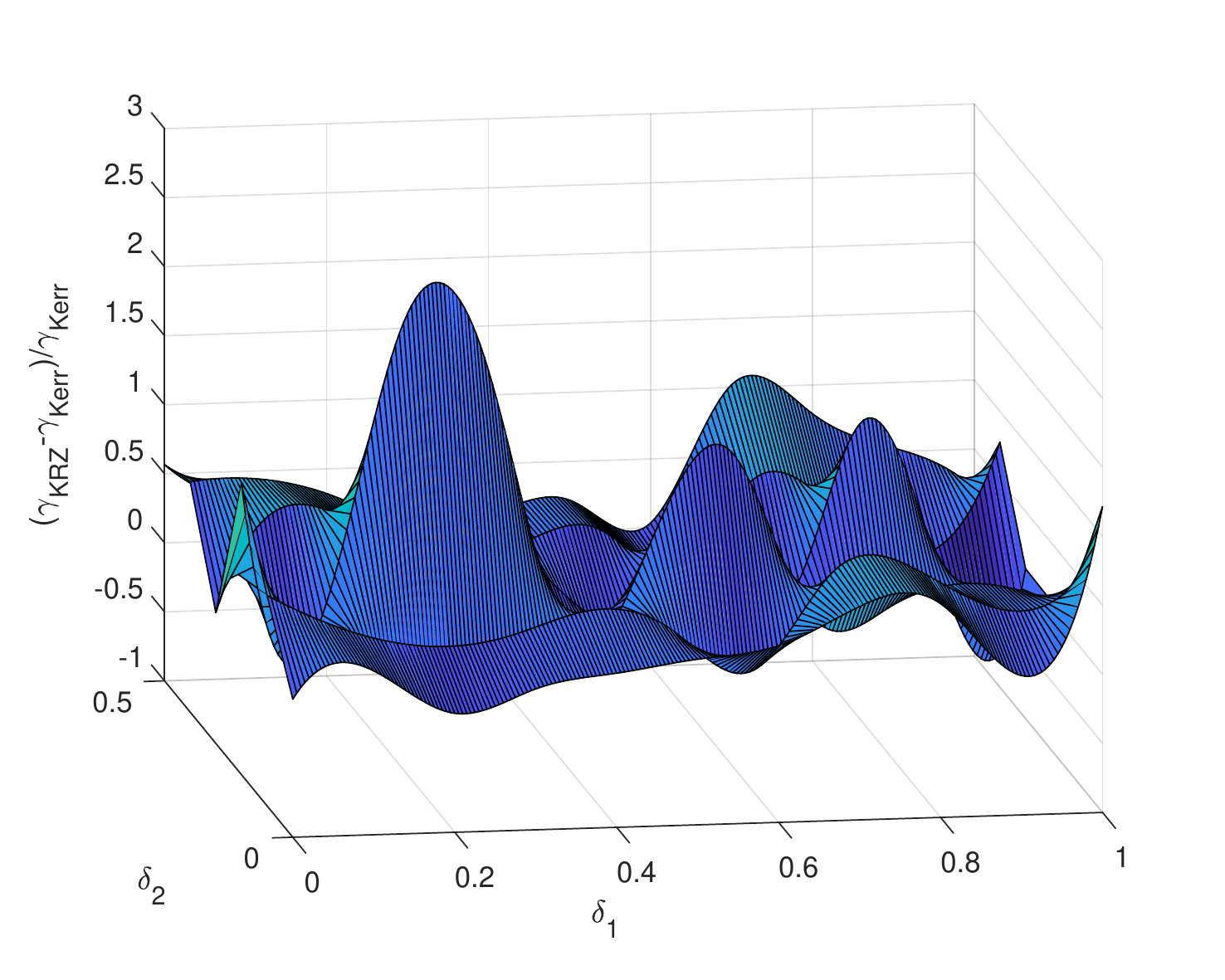}
		\includegraphics[width=0.40 \textwidth]{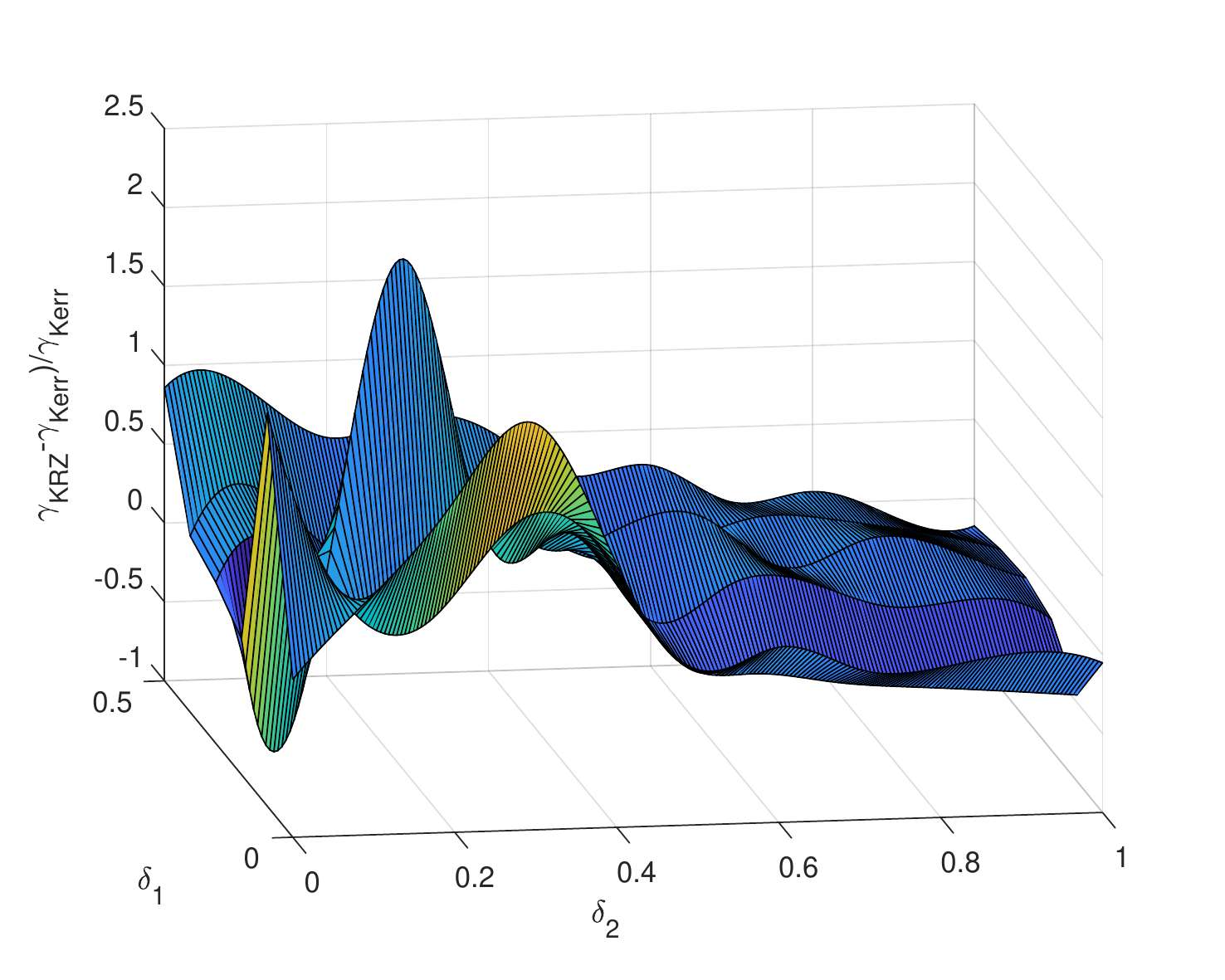}
\caption{Relative differences of LEs between the KRZ and Kerr cases as functions of $\delta_1$ and $\delta_2$. Left panel corresponds to the dependence from $\delta_{1}$ ranging from $0$ to $1$ and $\delta_{2}$ ranging from $0$ to $0.5$. Right panel correspond to the dependence from $\delta_{2}$ ranging from $0$ to $1$ and $\delta_{1}$ ranging from $0$ to $0.5$. Both panels can be regarded as 3D figures observed in different directions but the values of the coordinate axes are different. }\label{LE_fig_3D}
\end{figure*}

We have chosen the different orbits having separation in the $\dot{r}$ of the order $10^{-5}$ with the other  initial conditions to be the same. In Fig.~\ref{LE_fig_3D} we present the 3-D dependence of LE from the parameters $\delta_1$ and $\delta_2$.  From the Fig.~\ref{LE_fig_3D} one can easily see that when $\delta_{1}$ increases from 0 to 1, LE will have two maximum values and there is no obvious downward trend in the overall dependence. When $\delta_{2}$ increases from 0 to 1 LE  have a single maximum value and the overall trend of the graph is down.  From the results of Fig.~\ref{LE_fig_3D} one may speculate that this is related to the initial conditions: 
the $\dot{t_0}$ and $\dot{\phi_0}$ decrease with increasing $\delta_{1}$, the $\dot{t_0}$ and $\dot{\phi_0}$ increase with increasing $\delta_{2}$. Moreover, the angular momentum $l$ increases with the decrease of $\delta_{1}$ and $\delta_{2}$. However, the angular momentum $l$ decreases faster with the increase of $\delta_{2}$. Finally, we conclude that the results show in Fig.~\ref{LE_fig_3D} is related to the interaction of these initial conditions affected by the parameters of KRZ metric. 

From Fig.~\ref{LE_fig_3D}, we can find that the magnitude of LE is very sensitive to the KRZ parameters $\delta_1$ and $\delta_2$. Comparing with the frequency changes due to non-Kerr parameters, the decay rate of ringdown can constrain $\delta_1$ and $\delta_2$ better. However, from the figure, the variation of LE is very complicated and hard to be fitted. This creates a problem to construct a parameterized ringdown waveform.

\section{Ray-tracing code for photons}\label{rtcs}

The information from the the distant source in the KRZ spacetime comes through the study and analyze of the light ray from them to the observer. Through the light ray one can get an image of the source and consequently get some information. In this work we study the the trajectories of photons in the KRZ spacetime using the ray-tracing code described in the Ref.~\cite{Ray_code}. The code describes the trajectories of photons near the black hole. 

The evolution of the photon's position with different components: the $t$- and $\phi$-components can be obtained by the first-order differential equations~(\ref{conserved_E})-(\ref{conserved_L}). Then one can rewrite $p_t$ and $p_\phi$ in terms of two parameters: the normalized affine parameter $\lambda'=E/\lambda$  

\begin{equation}\label{dt}
\frac{d t}{d \lambda^{\prime}}=\frac{-g_{\phi \phi}-b g_{t \phi}}{g_{\phi \phi} g_{t t}-g_{t \phi}^{2}}\ ,
\end{equation}
\begin{equation}\label{dphi}
\frac{d \phi}{d \lambda^{\prime}}=\frac{b g_{t t}+b g_{t \phi}}{g_{\phi \phi} g_{t t}-g_{t \phi}^{2}}\ .
\end{equation}

For the remaining $r$- and $\phi$- components of the photon's position in the KRZ space-time, we can use the second-order geodesic equations with the normalized affine parameter and the Christoffel symbols $\Gamma^\sigma_{\mu\nu}$ as:

\begin{equation}\label{geod}
\frac{d^2x^\sigma}{d\lambda'^2}+\Gamma^\sigma_{\mu\nu}\frac{dx^\mu}{d\lambda '} \frac{dx^\nu}{d\lambda'}=0. \\
\end{equation}

In this way we can get the system of equations that the ray-tracing code can be used for KRZ spacetime. 

We suppose that the massive source described by the KRZ spacetime is located at the origin of the reference frame and coordinate system. We choose the mass of the object $M=1$ since it does not affect the shape of the shadow. We assume that the observer's screen is located at a distance away the source of $d=1000$, the azimuthal and polar angles are $\gamma_{rt}$ and $0$, respectively. The celestial coordinates $(\alpha, \beta)$ on the observer's sky are related to polar coordinates $r_{\rm scr}$ and $\phi_{\rm scr}$ on the screen by $\alpha=r_{\rm scr}\cos(\phi_{\rm scr})$ and $\beta=r_{\rm scr}\sin(\phi_{\rm scr})$. Since we only know the positions and momenta of the photon in the screen, we should solve the geodesic equations from the screen to the source. The photons depart from the screen with a four-momentum perpendicular to the screen and other initial conditions. The method assumes that the screen at spatial infinity, only the photons which moving perpendicular to the screen at a distance $d$ could influence the infinite screen.

The initial position and four-momentum of each photon in the KRZ spacetime are given as~\cite{source_screen}
\begin{equation}\label{in_pos_r}
		 r_i =\left(d^2+\alpha^2+\beta^2\right)^{1/2}      ,\\
\end{equation}
\begin{equation}\label{in_pos_theta}
		 \theta_i=\arccos\left(\frac{d\cos\gamma_{rt}+\beta\sin\gamma_{rt}}{r_i}\right),\\
\end{equation}
\begin{equation}\label{in_pos_phi}
		 \phi_i = \arctan\left(\frac{\alpha}{d\sin\gamma_{rt}-\beta\cos\gamma_{rt}}\right),
\end{equation}
and
\begin{equation}\label{in_momen_r}
	 \left(\frac{dr}{d\lambda'}\right)_i=\frac{d}{r_i},\\
\end{equation}
\begin{equation}\label{in_momen_theta}
	 \left(\frac{d\theta}{d\lambda'}\right)_i=\frac{-\cos\gamma_{rt}+\frac{d}{r_i^2}(d\cos\gamma_{rt}+\beta\sin\gamma_{rt})}{\sqrt{r_i^2-(d\cos\gamma_{rt}+\beta\sin\gamma_{rt})^2}},\\
\end{equation}
\begin{equation}\label{in_momen_phi}
	 \left(\frac{d\phi}{d\lambda'}\right)_i=\frac{-\alpha\sin\gamma_{rt}}{\alpha^2+(d\cos\gamma_{rt}+\beta\sin\gamma_{rt})^2},\\
\end{equation}

Using of the equation~(\ref{LRs}) of the photon four-momentum to be zero one can find the component $(dt/d\lambda')_i$. The conserved quantity $b$, which is involved in Eqs.~(\ref{dt}) and (\ref{dphi}), is calculated from the initial conditions of $E$ and $L_{z}$.

The initial conditions of the code on the screen is defined in the following way. The confines of the location of the compact source is found inside $0 \leq r_{\rm scr} \leq 20$, and the value of the $\phi_{\rm scr}$  in the range $0 \leq \phi_{\rm scr} \leq 2\pi$ with step of $\pi / 180$. The confines is the border between the photons that are captured by the compact source and the photons that are able to escape to  infinity. The photons are considered as captured by the compact source if they cross the surface $r = r_{\rm surf} + \delta r$ with $\delta r = 10^{-3}$, where $r_{\rm surf}$ is the radius of the horizon. Then the confines is amplified in to an accuracy of $\delta r_{\rm scr} = 10^{-3}$ to accurately determine the shadow boundary with the value of $r_{\rm scr}$ for the corresponding value of $\phi_{\rm scr}$. This method allows one to accurately calculate the shadow produced by light ray in the KRZ parametrized metric with high accuracy with respect to sampling the entire screen. 

\section{The shadow of the KRZ metric \label{shadow}}

\begin{figure}
	\includegraphics[width=0.5 \textwidth]{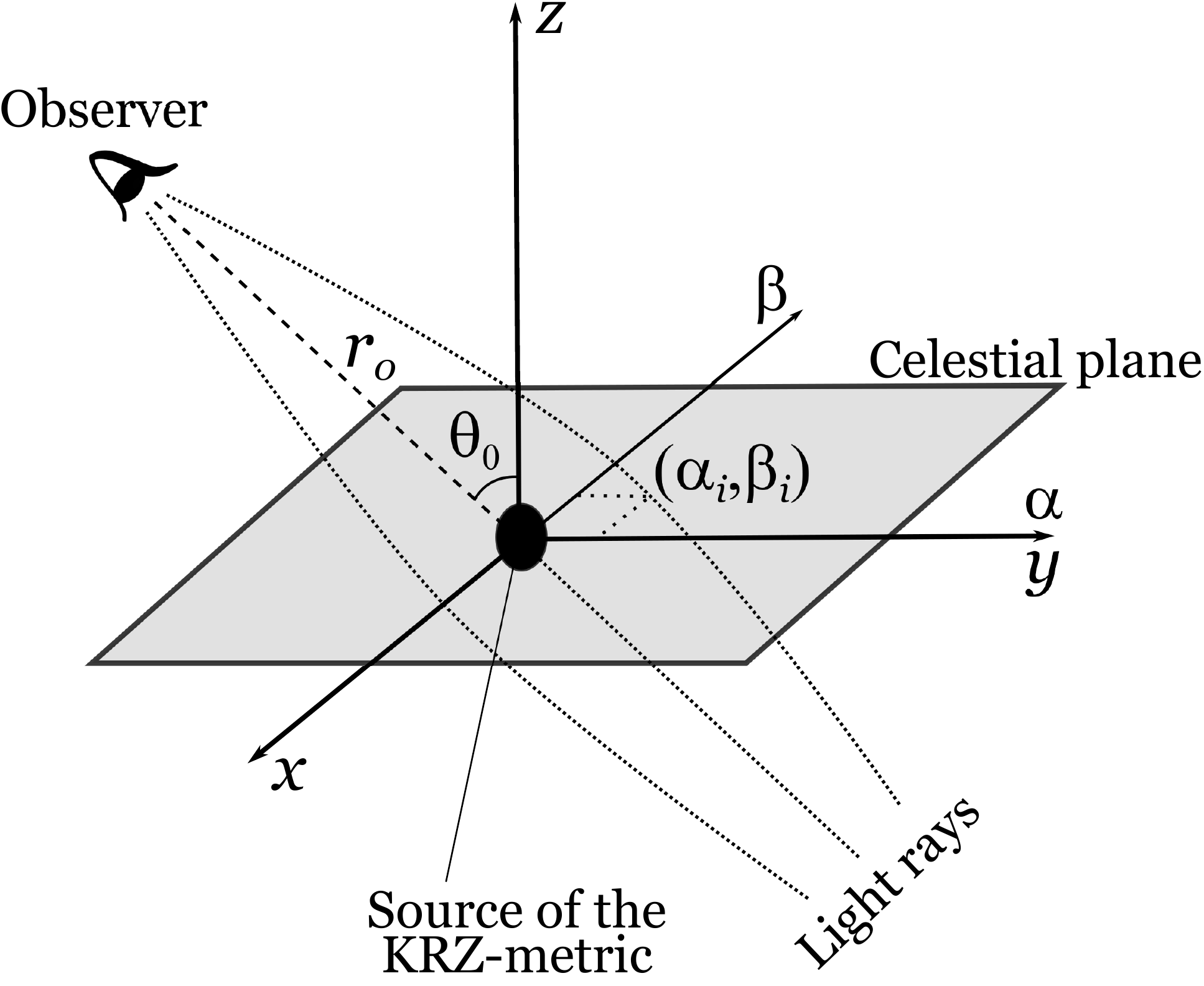}

	\caption{Schematic illustration of the celestial coordinates used for the ray tracing code in the KRZ spacetime. \label{shadsch}}
\end{figure}

In this section, we plan to study the apparent shape of the compact object shadow under the KRZ spacetime.
We can use the celestial coordinates $\alpha$ and $\beta$~\cite{Cele_coor_1} to describe the shadow of the compact object described by the KRZ spacetime
\begin{eqnarray}
\alpha &=&\underset{r_0 \rightarrow \infty}{\lim}\left(-r_0^2 \sin \theta_0 \frac{d\phi}{dr}\right) \ , \label{eq:14}\\
\beta &=&\underset{r_0 \rightarrow \infty}{\lim}\left(r_0^2 \frac{d\theta}{dr}\right) \ , \label{eq:15}
\end{eqnarray}
where $r_0$ is the distance between the massive source and observer and $\theta_0$ is the inclination angle between the observer lens axis and the normal of observer's sky plane~(see Fig.~\ref{shadsch}). 

\begin{figure*}
\centering
\subfigure[\;a-Kerr]{
\includegraphics[width=0.4 \textwidth]{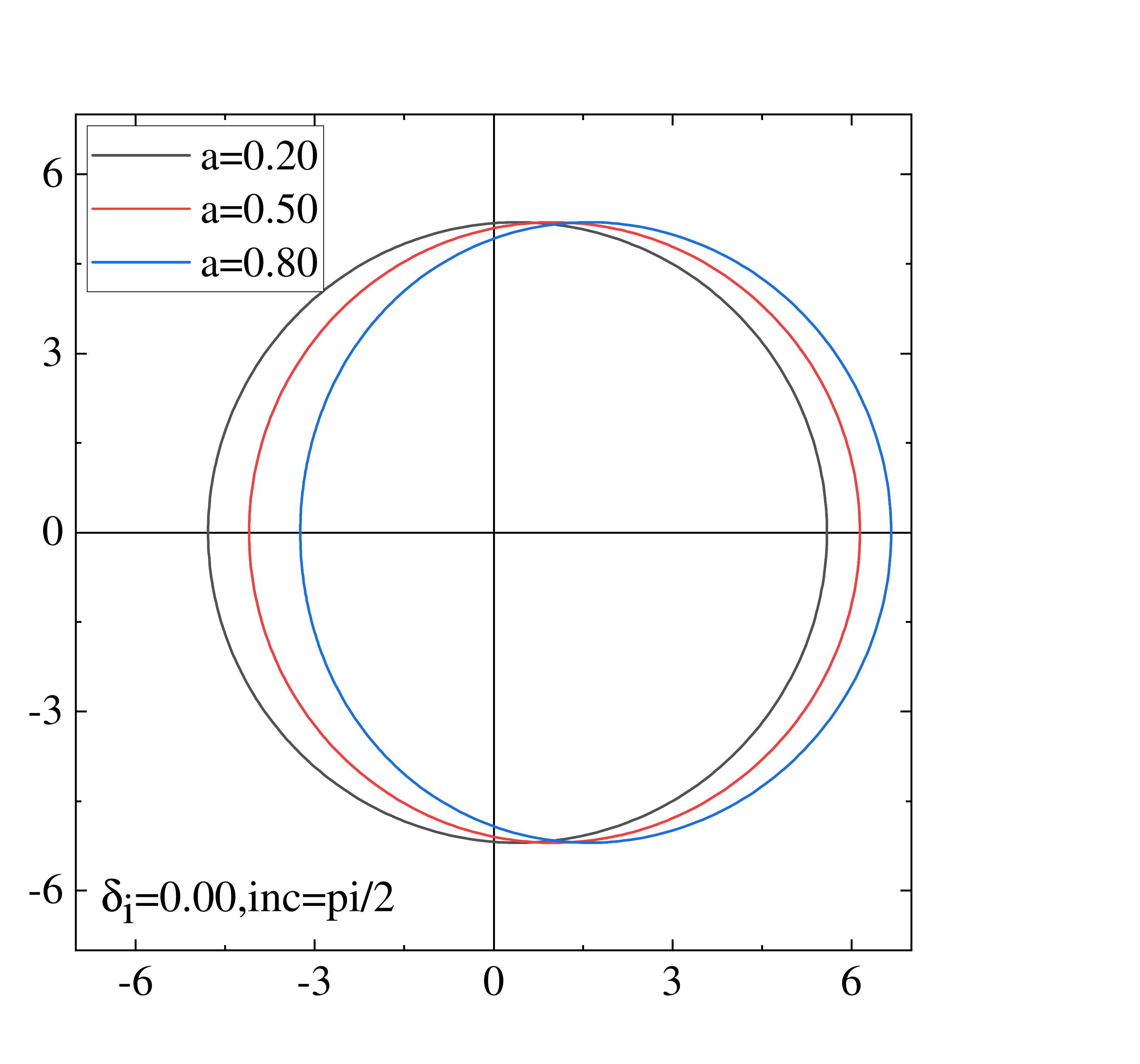}
}
\quad
\subfigure[\;a]{
\includegraphics[width=0.4 \textwidth]{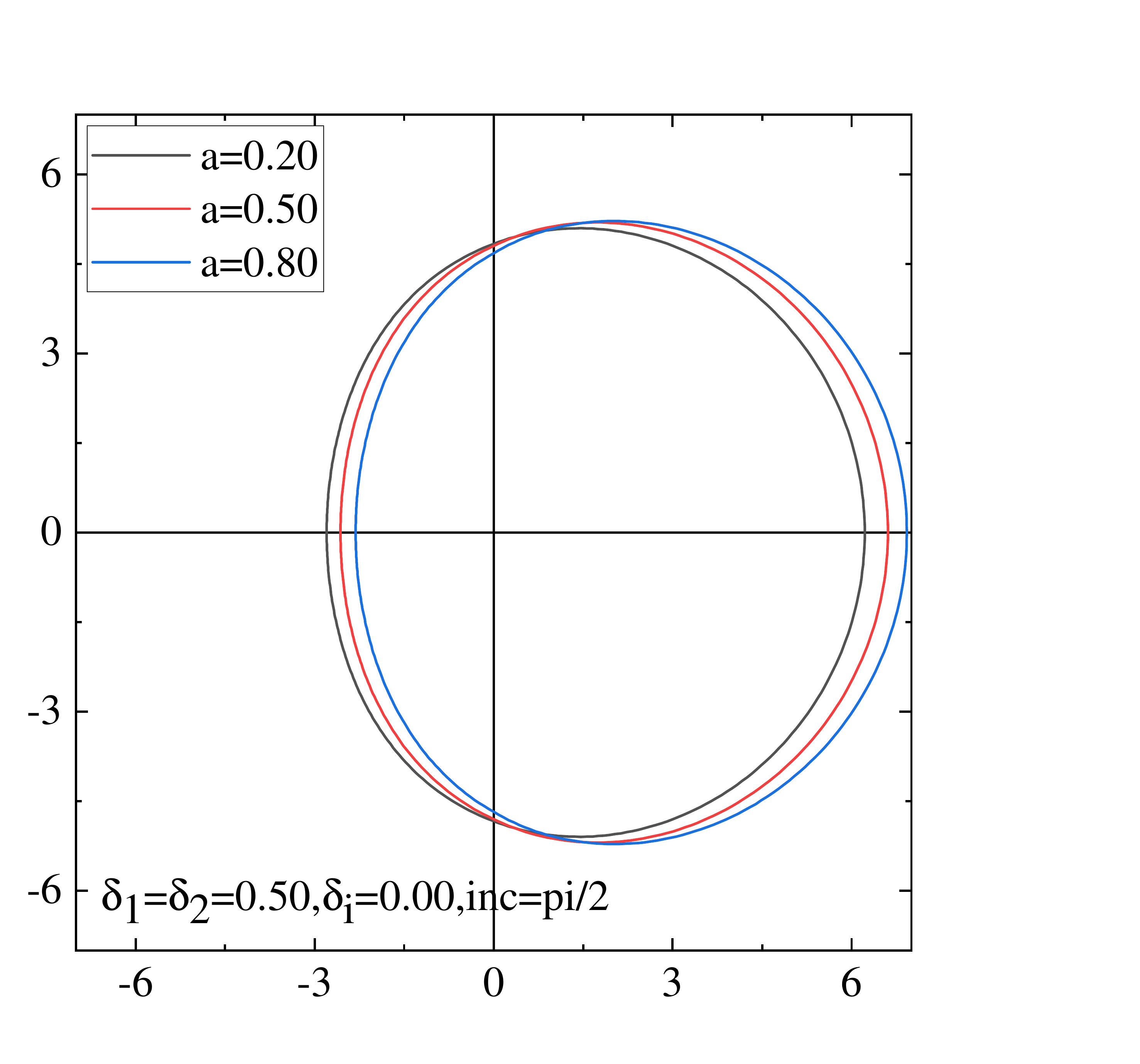}
}
\quad
\subfigure[\;inclination-Kerr]{
\includegraphics[width=0.4 \textwidth]{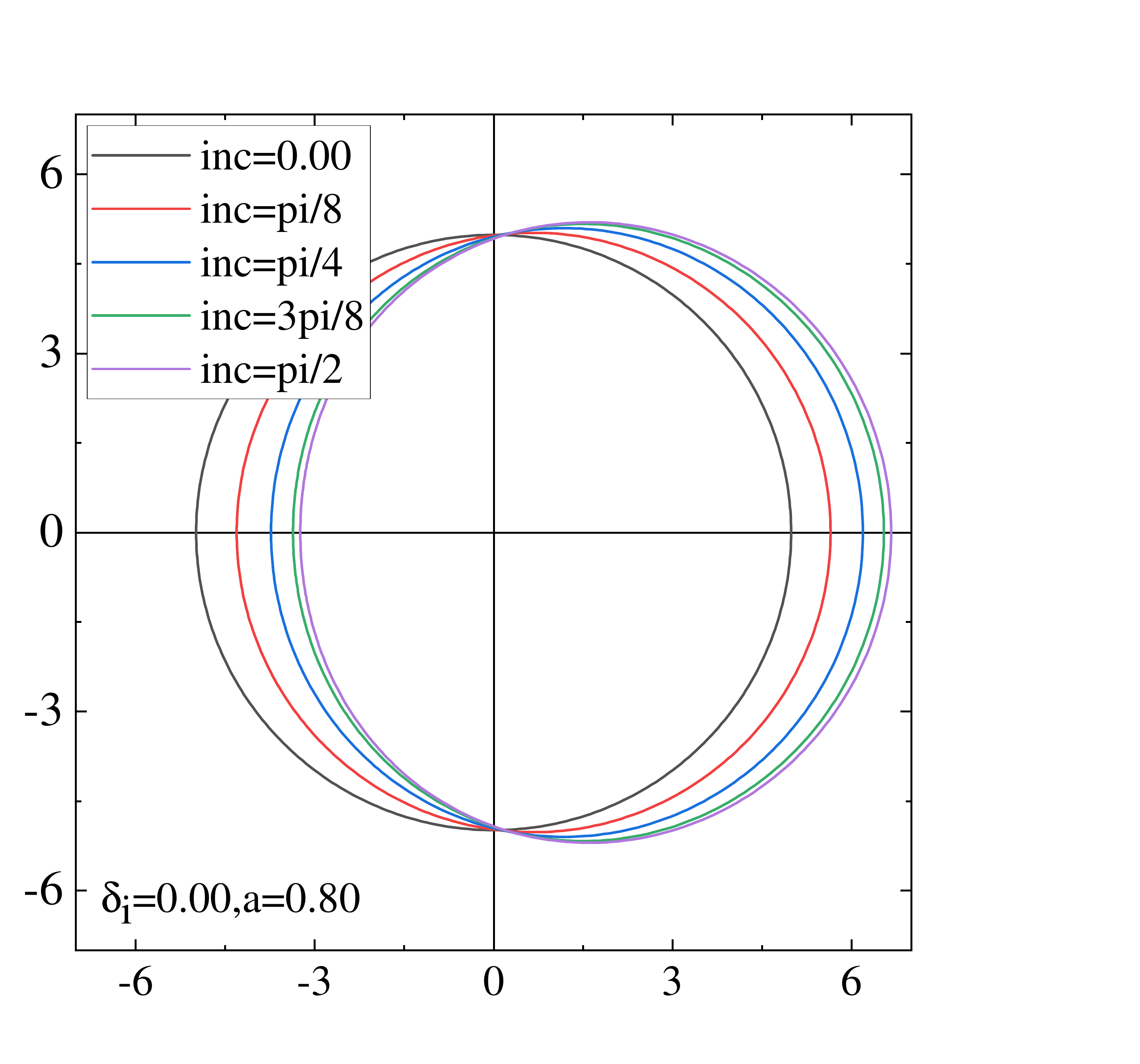}
}
\quad
\subfigure[\;inclination]{
\includegraphics[width=0.4 \textwidth]{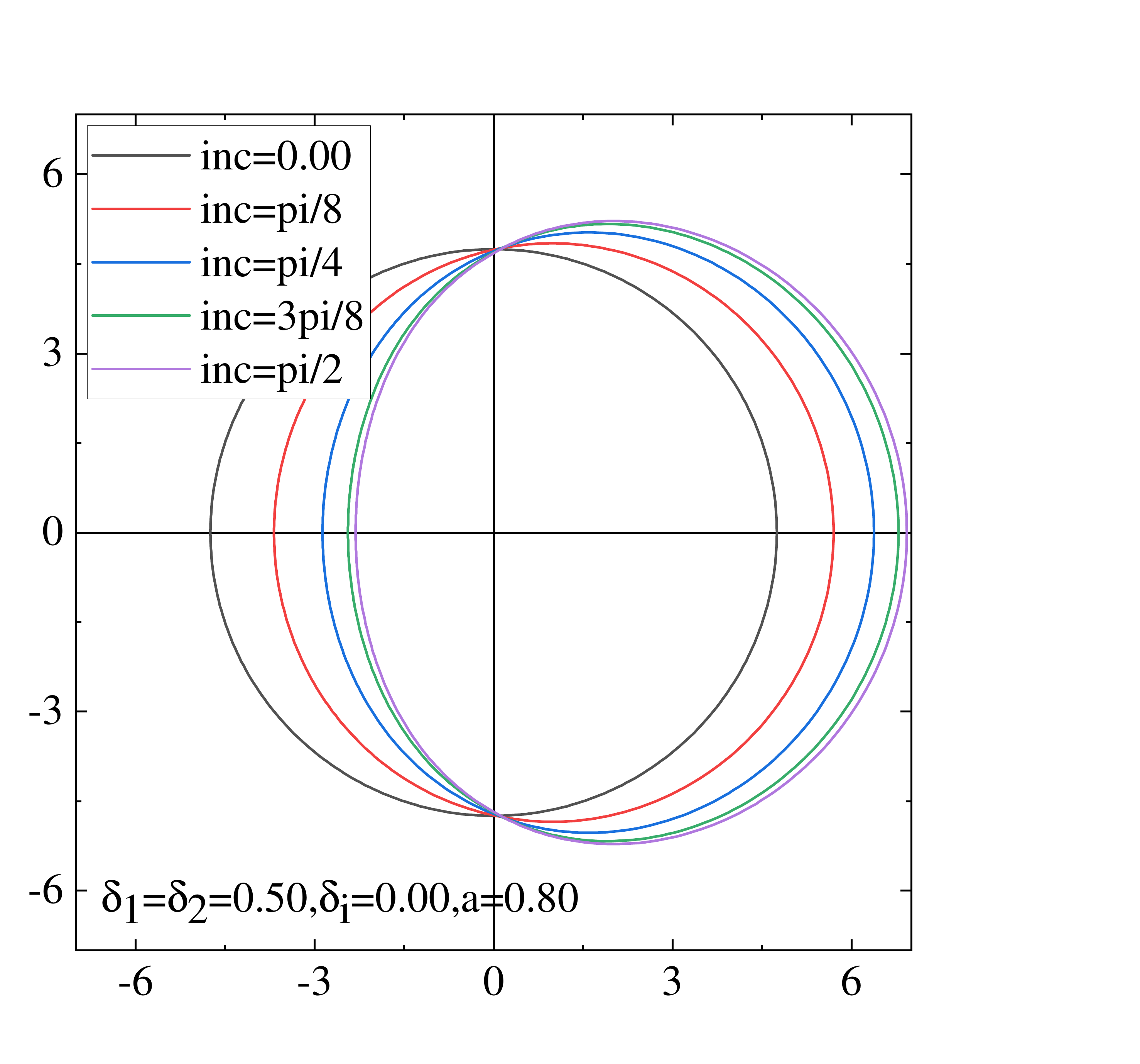}
}
\caption{Ray-traced shadow images in the Kerr and KRZ spacetime. The (a) is for the shadow with different spin $a$ in the Kerr spacetime, the (b) is for the shadow with different spin $a$ in the KRZ spacetime, the (c) is for the shadow with different inclination in the Kerr spacetime, the (d) is for the shadow with different inclination in the KRZ spacetime\label{shadow_Diff_metri}}
\end{figure*}

\begin{figure*}
\centering
\subfigure[\;$\delta_{1}$]{
\includegraphics[width=0.4 \textwidth]{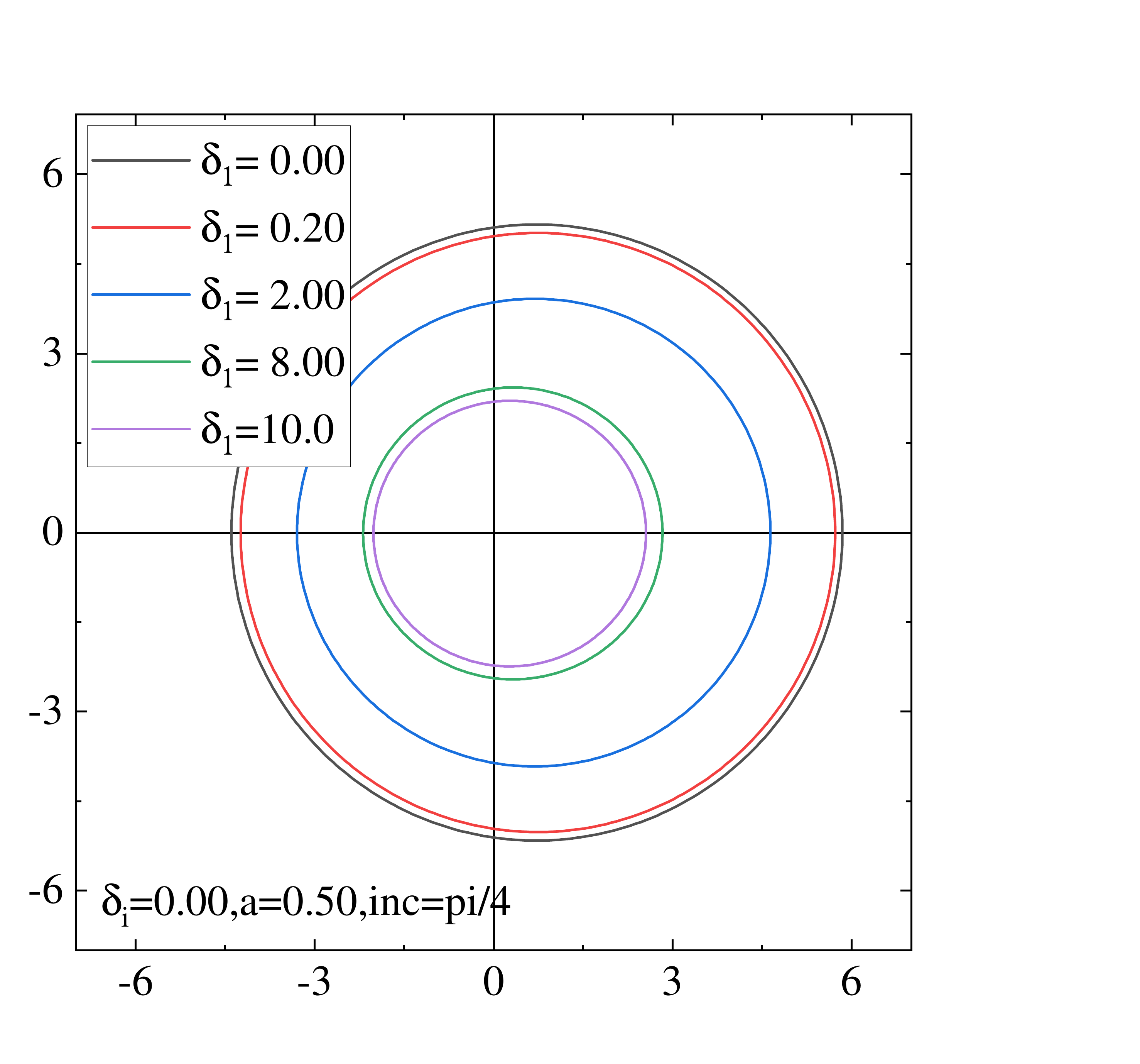}
}
\quad
\subfigure[\;$\delta_{2}$]{
\includegraphics[width=0.4 \textwidth]{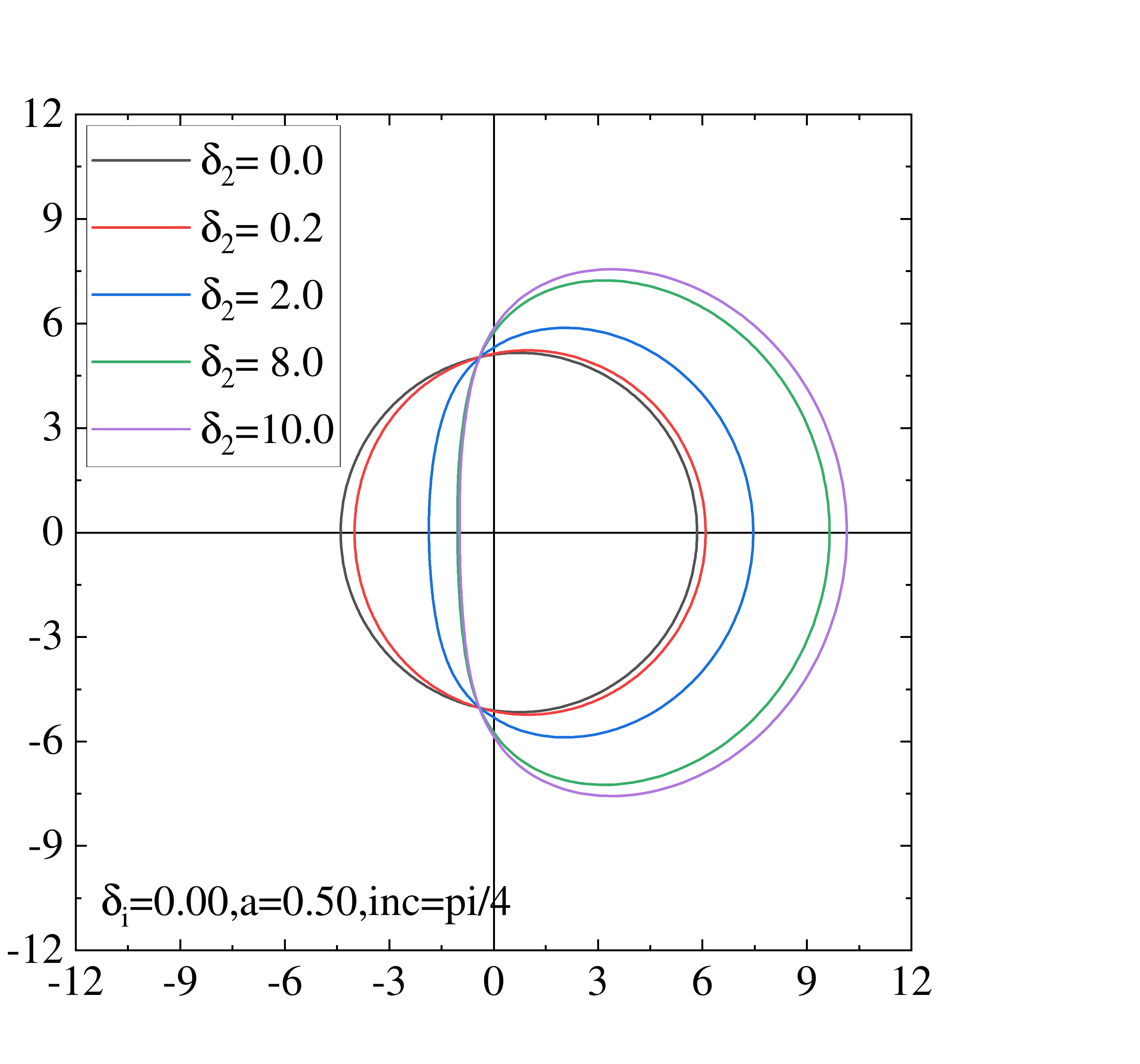}
}
\quad
\subfigure[\;$\delta_{3}$]{
\includegraphics[width=0.4 \textwidth]{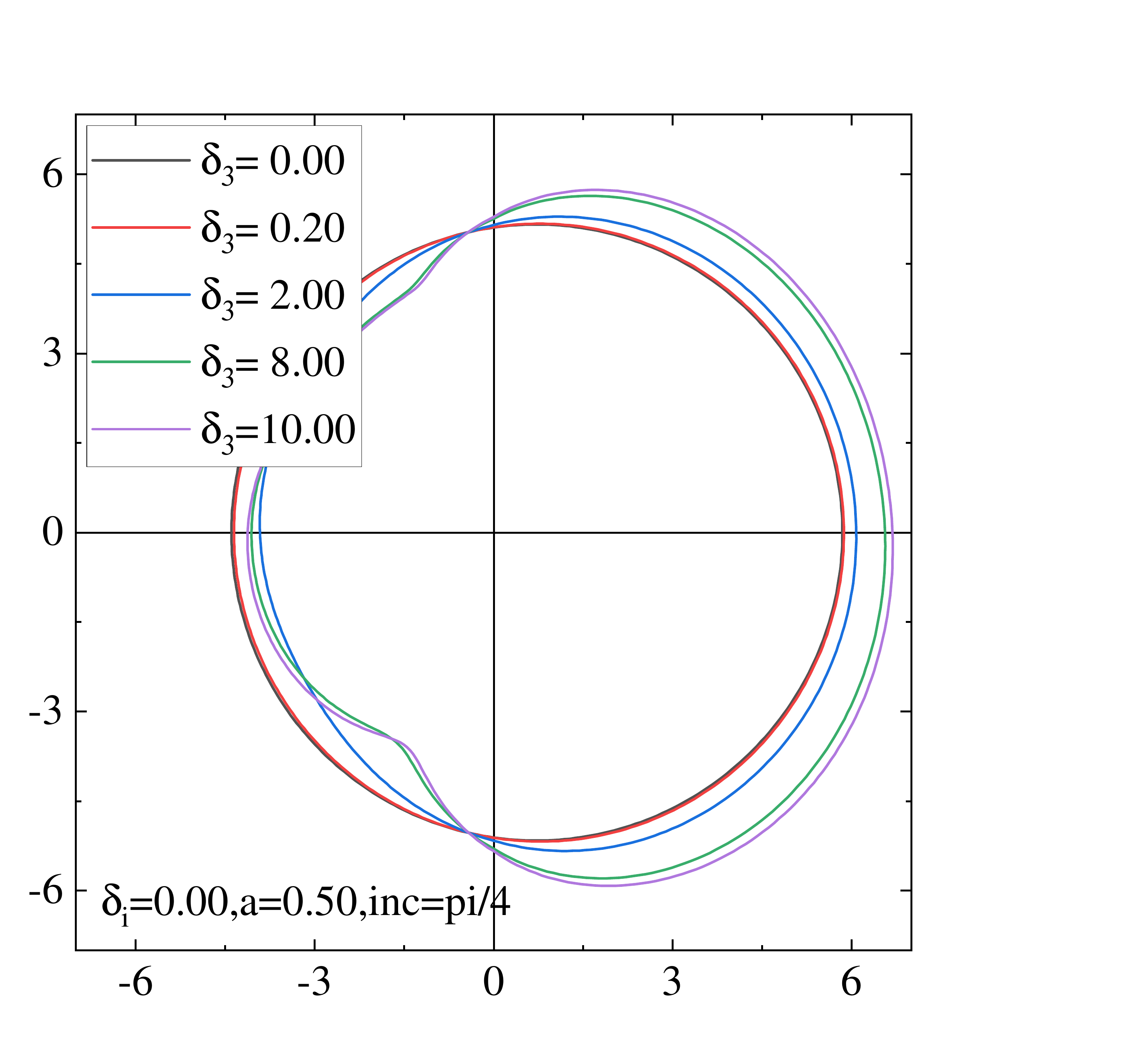}
}
\quad
\subfigure[\;$\delta_{4}$]{
\includegraphics[width=0.4 \textwidth]{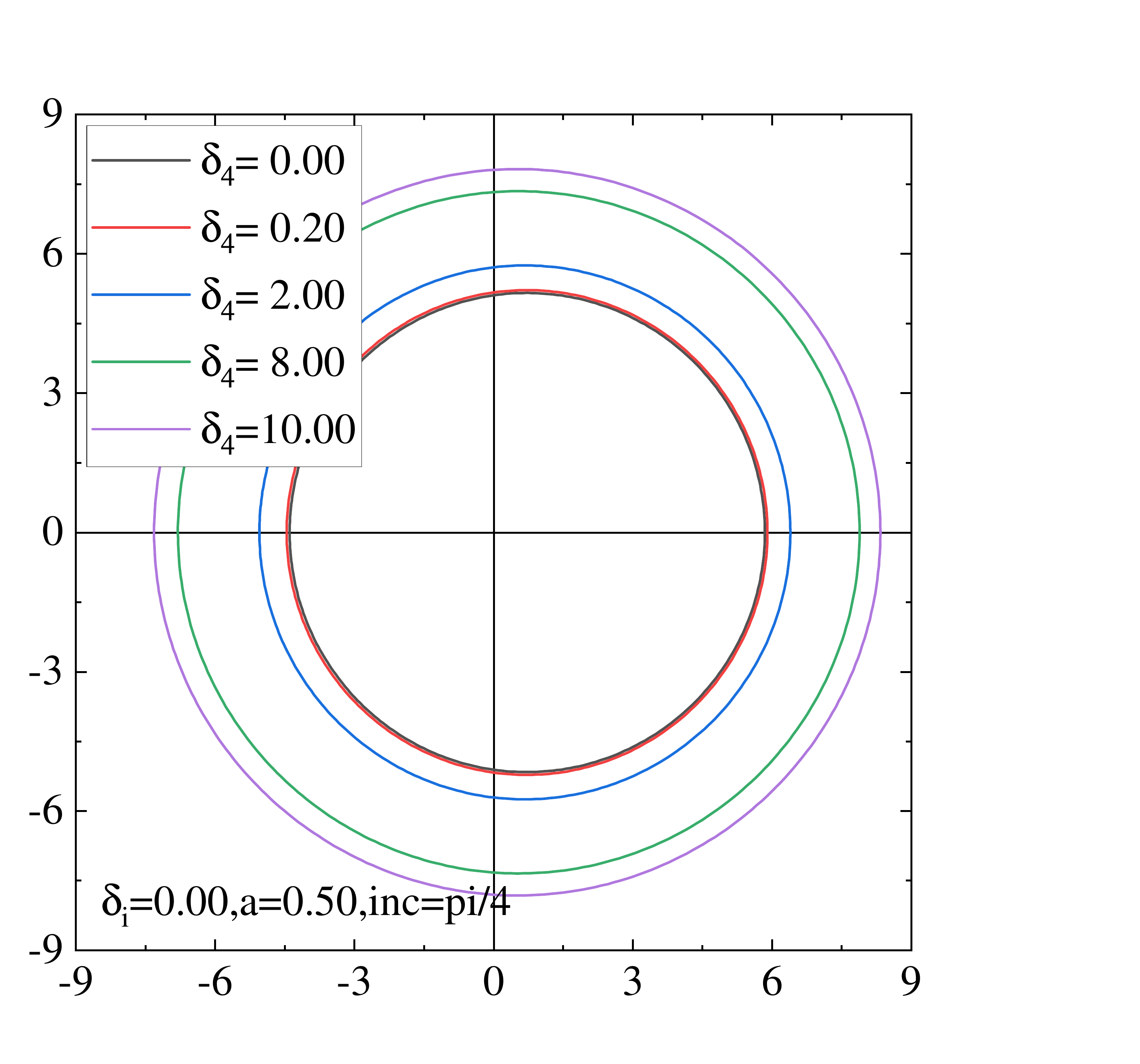}
}
\quad
\subfigure[\;$\delta_{5}$]{
\includegraphics[width=0.4 \textwidth]{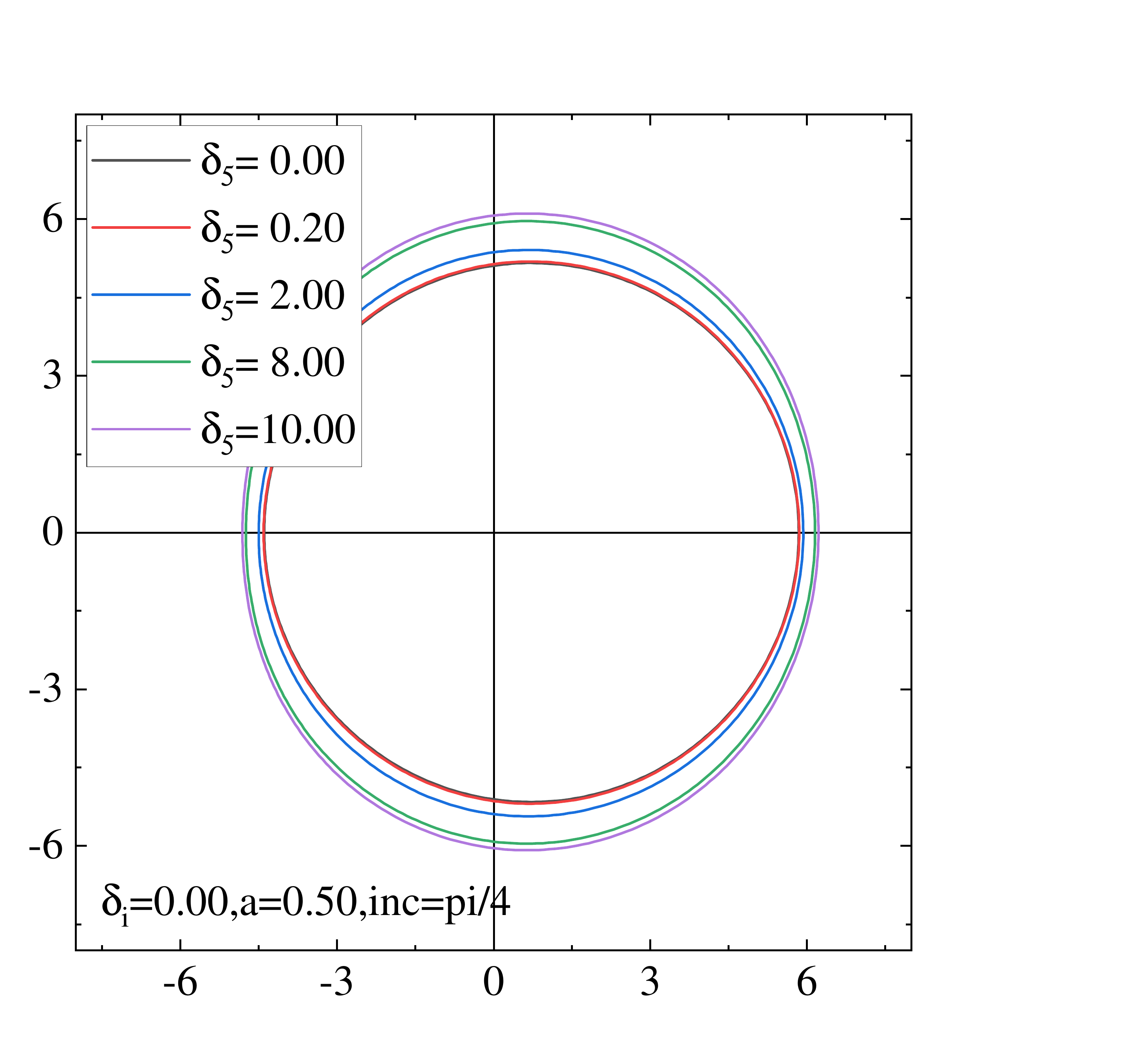}
}
\quad
\subfigure[\;$\delta_{6}$]{
\includegraphics[width=0.4 \textwidth]{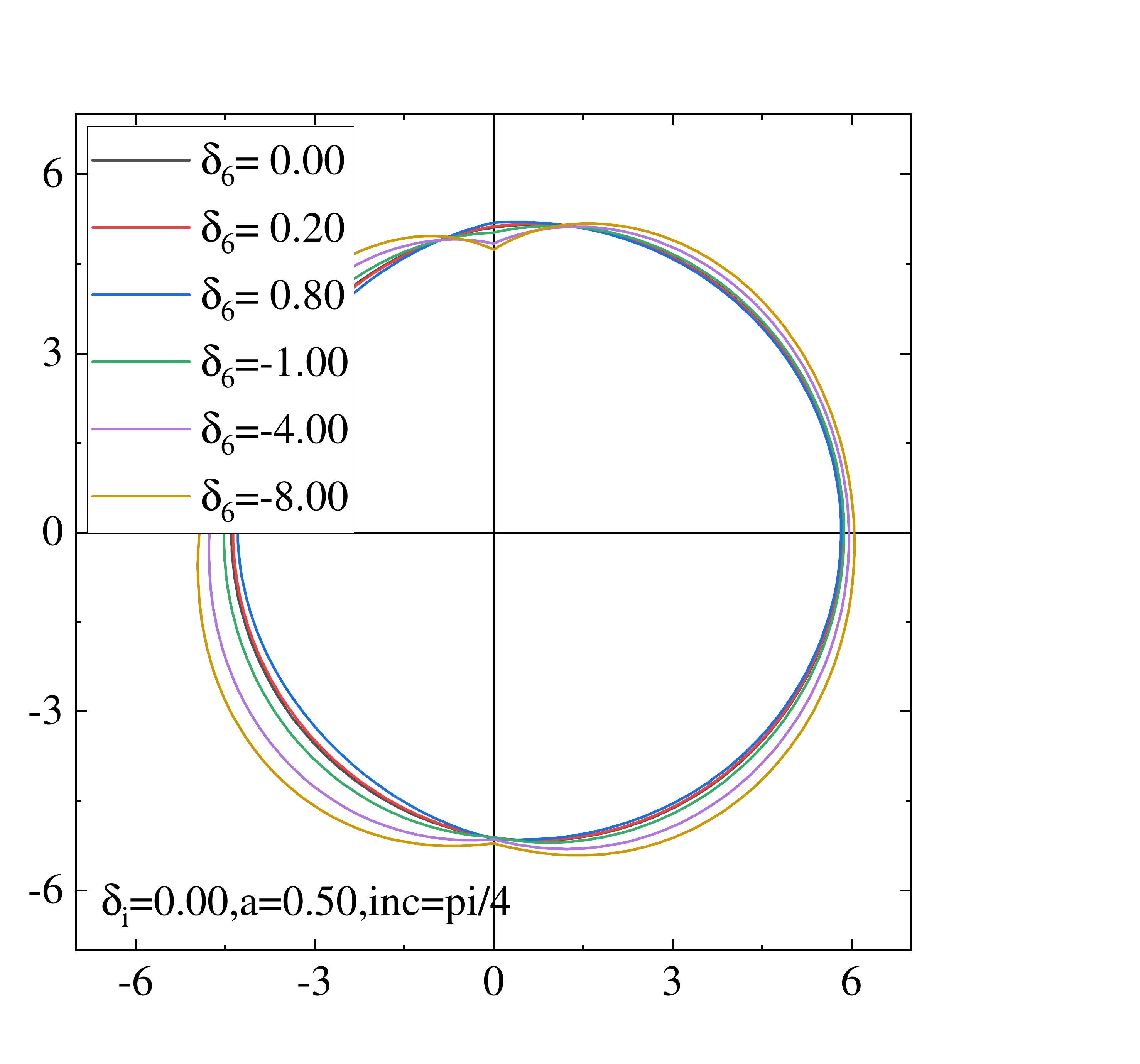}
}
\caption{Ray-traced shadow images in the KRZ spacetime. The (a)-(f) are for the shadow corresponding different values of $\delta$. And the text $\delta_{i}=0$ in the graph  means that other parameters except one varying in the plot equal to zero.\label{shadow_delta}}
\end{figure*}

\begin{figure}
	\includegraphics[width=0.5 \textwidth]{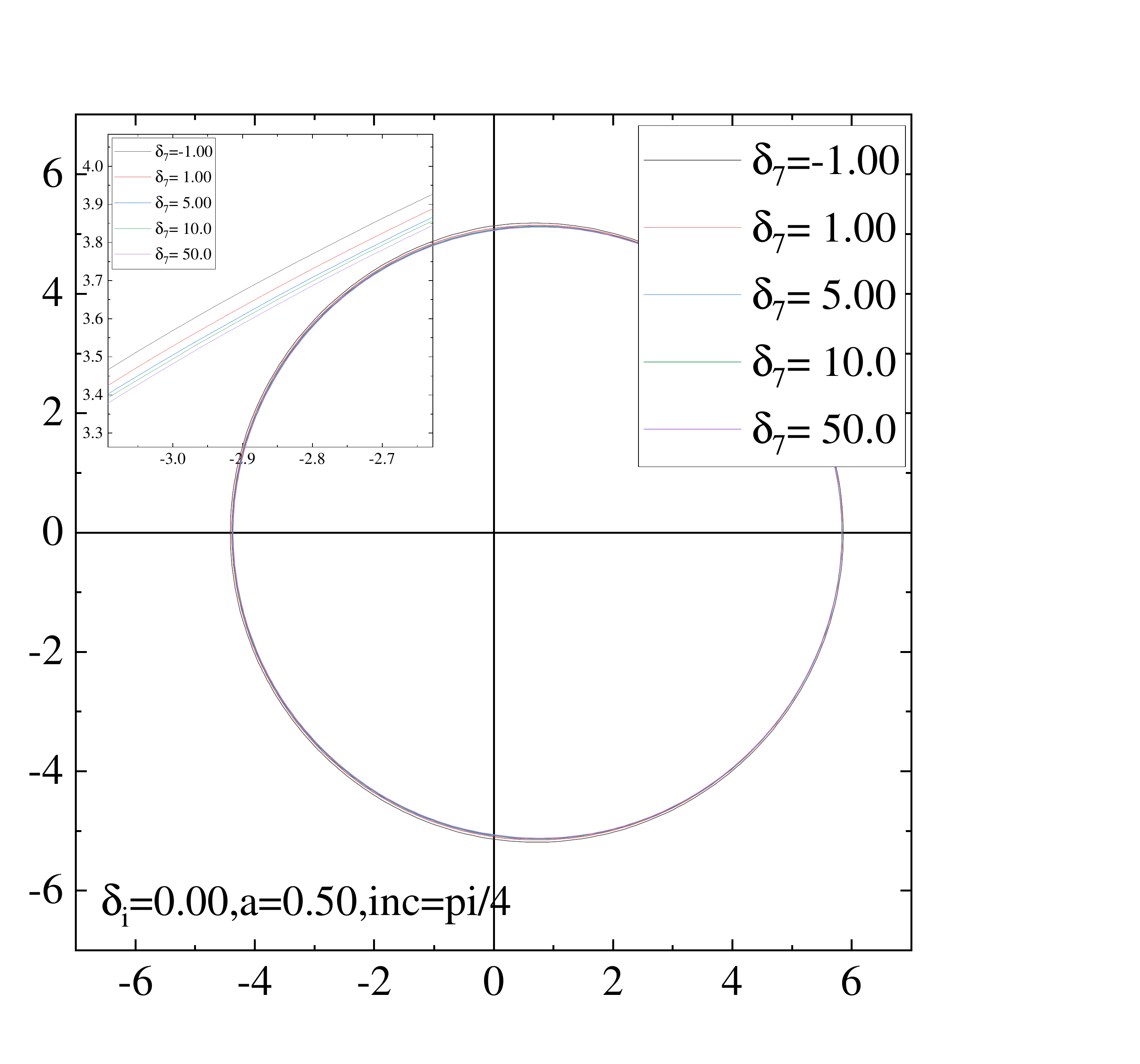}
	\caption{The shadow of the black hole for the different values of parameter $\delta_{7}$ in the KRZ spacetime. \label{delta7}}
\end{figure}

\begin{figure}
	\includegraphics[width=0.5 \textwidth]{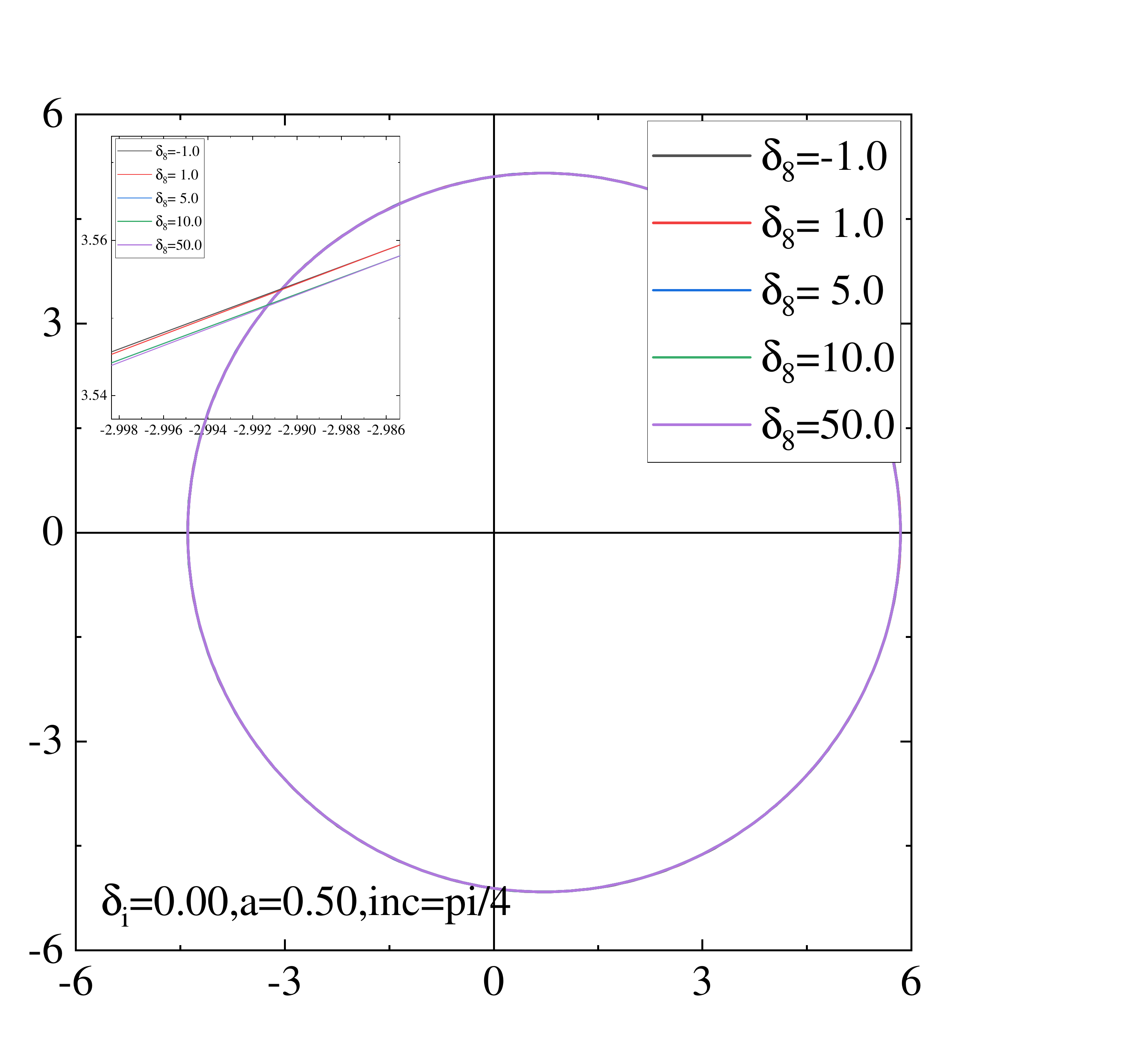}
	\caption{The shadow of the black hole for the different values of parameter $\delta_{8}$ in the KRZ spacetime. \label{delta8}}
\end{figure}

\begin{figure*}
\centering
\subfigure[\;]{
\includegraphics[width=0.4 \textwidth]{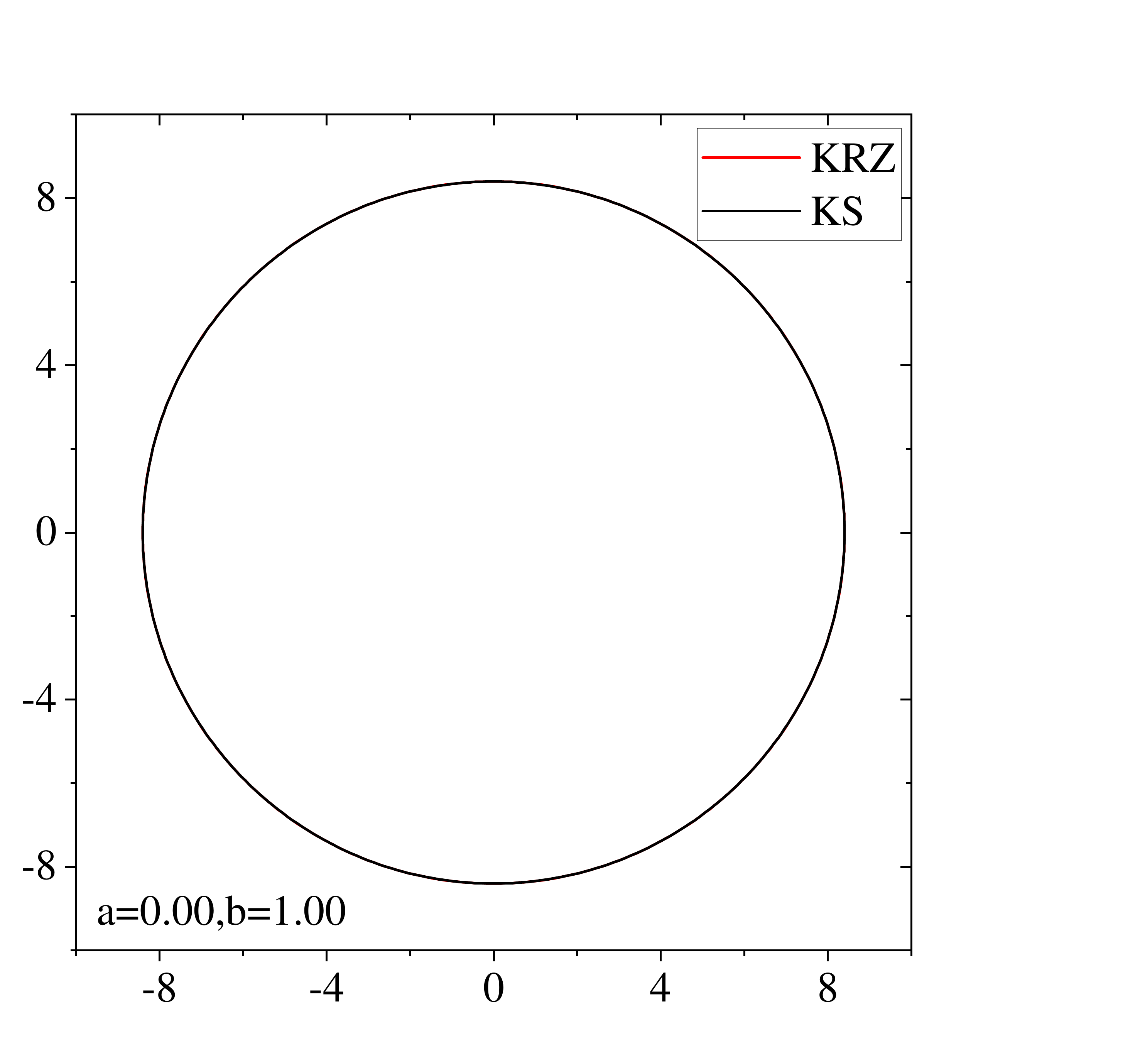}
}
\quad
\subfigure[\;]{
\includegraphics[width=0.4 \textwidth]{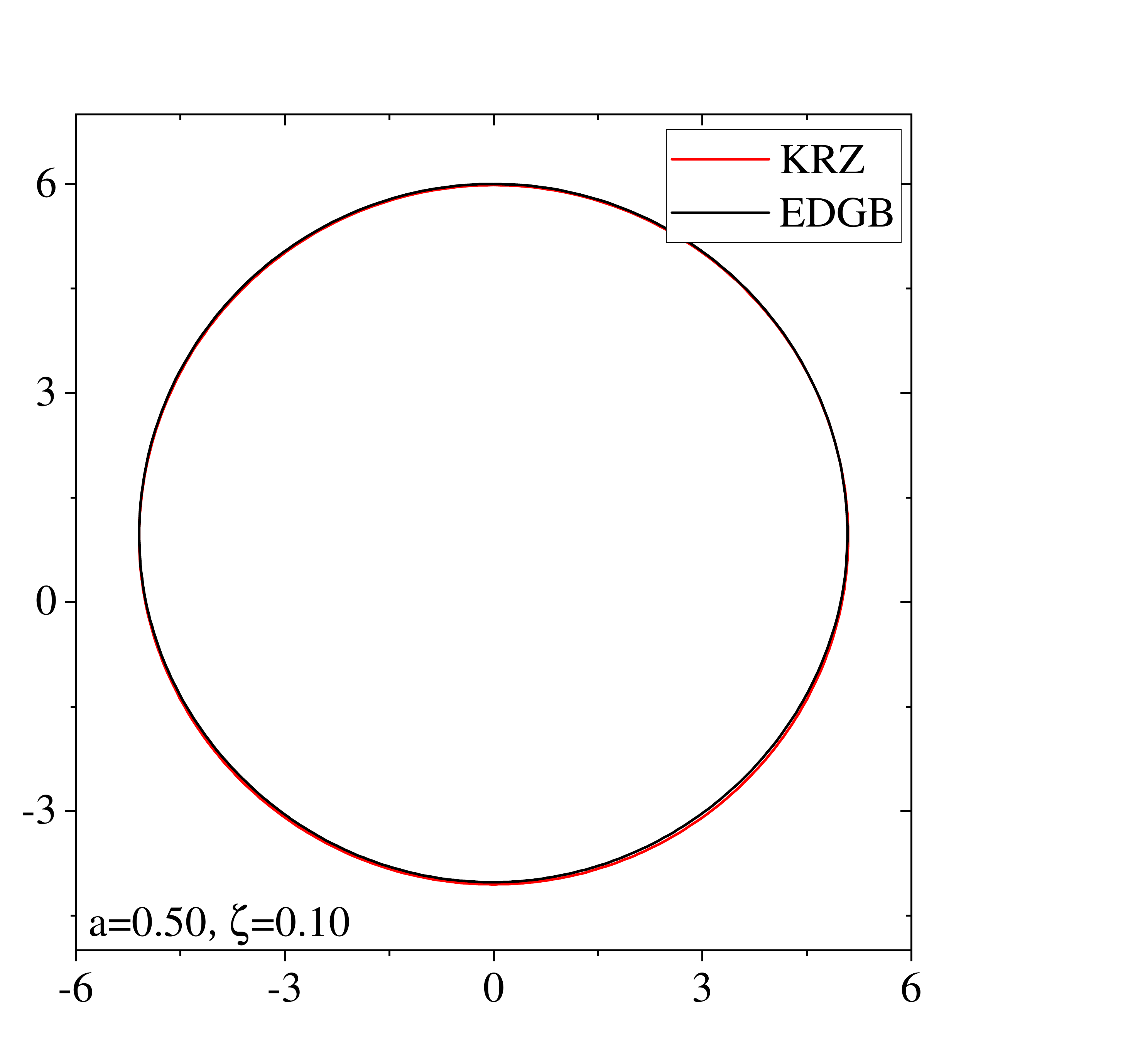}
}
\caption{The shdow of the black hole for the different metirc,in figure (a) we compare the KS metric and KRZ metric then in figure (b) we compare the EDGB metric and KRZ metric(a is the value of spin, b is the value of scalar(dilaton) field and $\zeta$ is the value of deformation)}.\label{shadow_ES_EDGB}
\end{figure*}

\begin{figure*}
\centering
\subfigure[\;$\langle R \rangle$-a]{
\includegraphics[width=0.40 \textwidth]{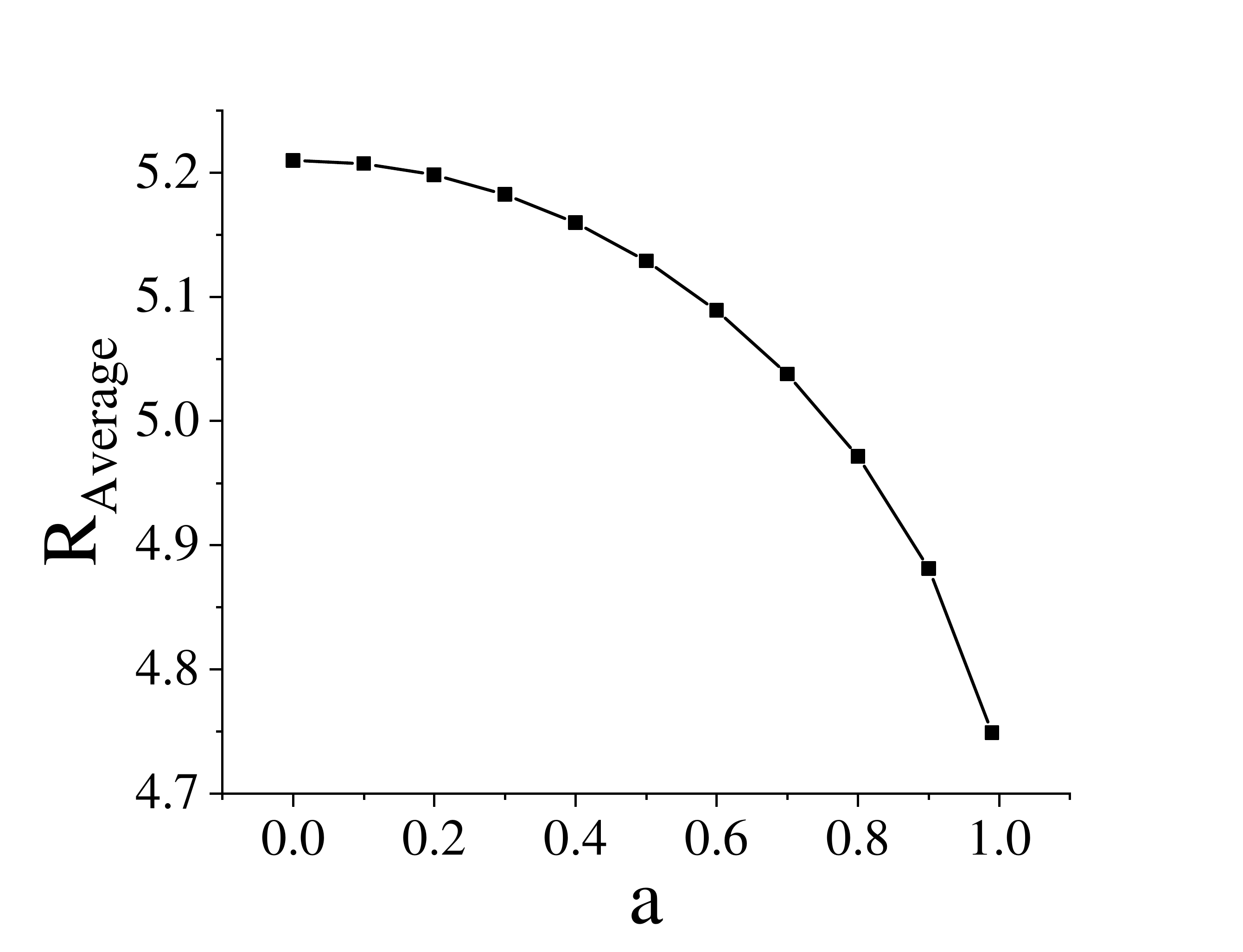}
}
\quad
\subfigure[\;$\langle R \rangle$-$\delta_{i}$]{
\includegraphics[width=0.40 \textwidth]{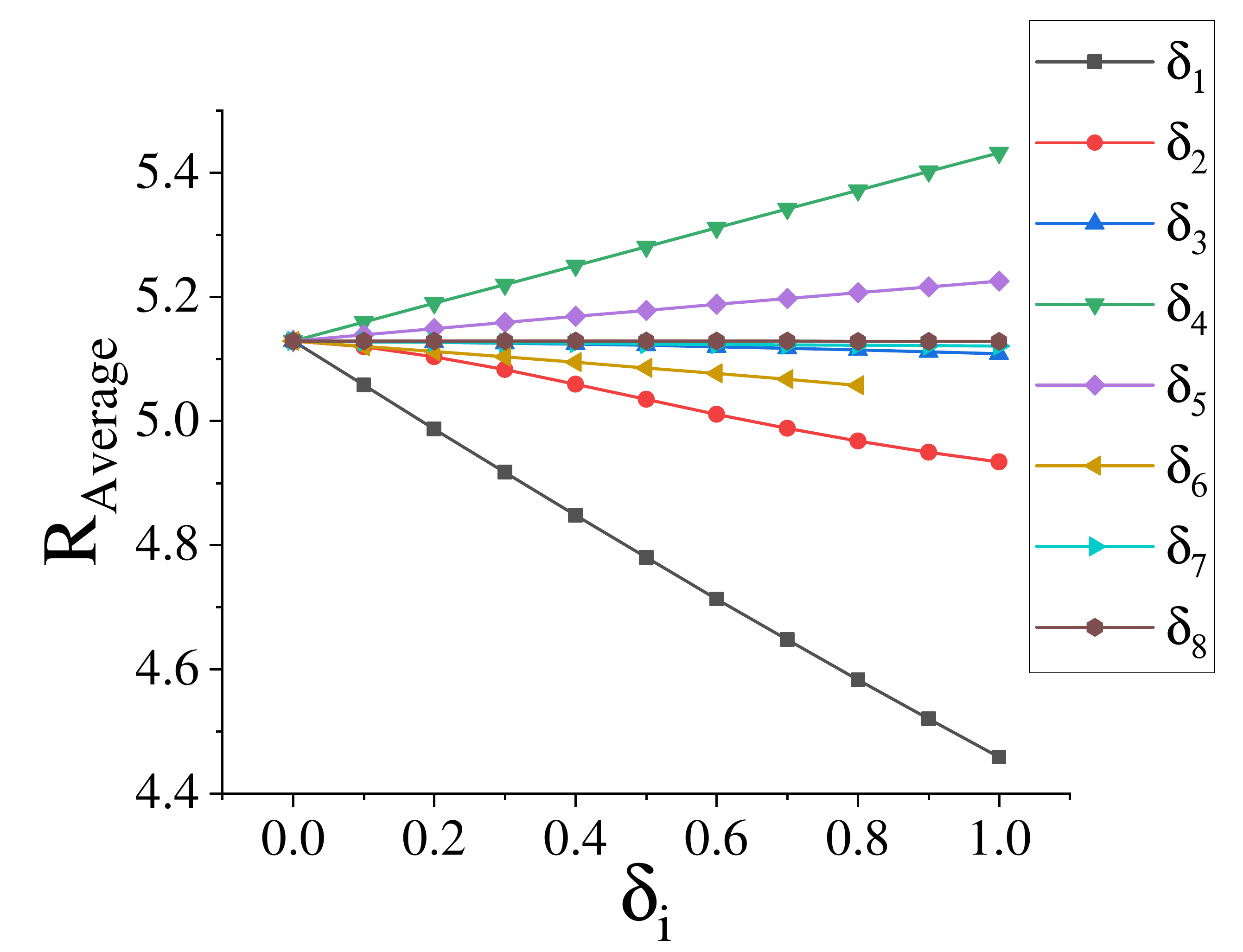}
}
\quad
\subfigure[\;A-a]{
\includegraphics[width=0.40 \textwidth]{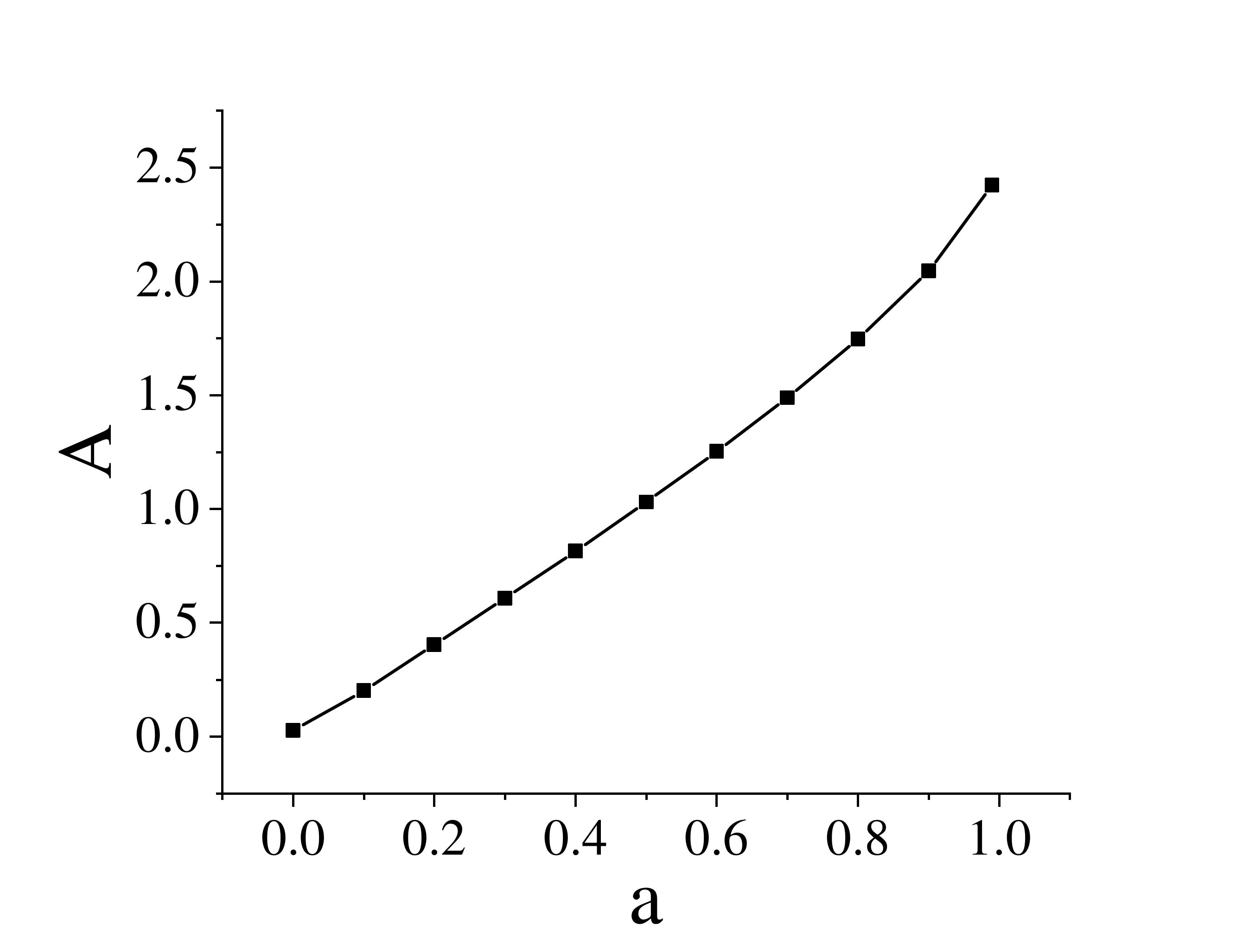}
}
\quad
\subfigure[\;A-$\delta_{i}$]{
\includegraphics[width=0.40 \textwidth]{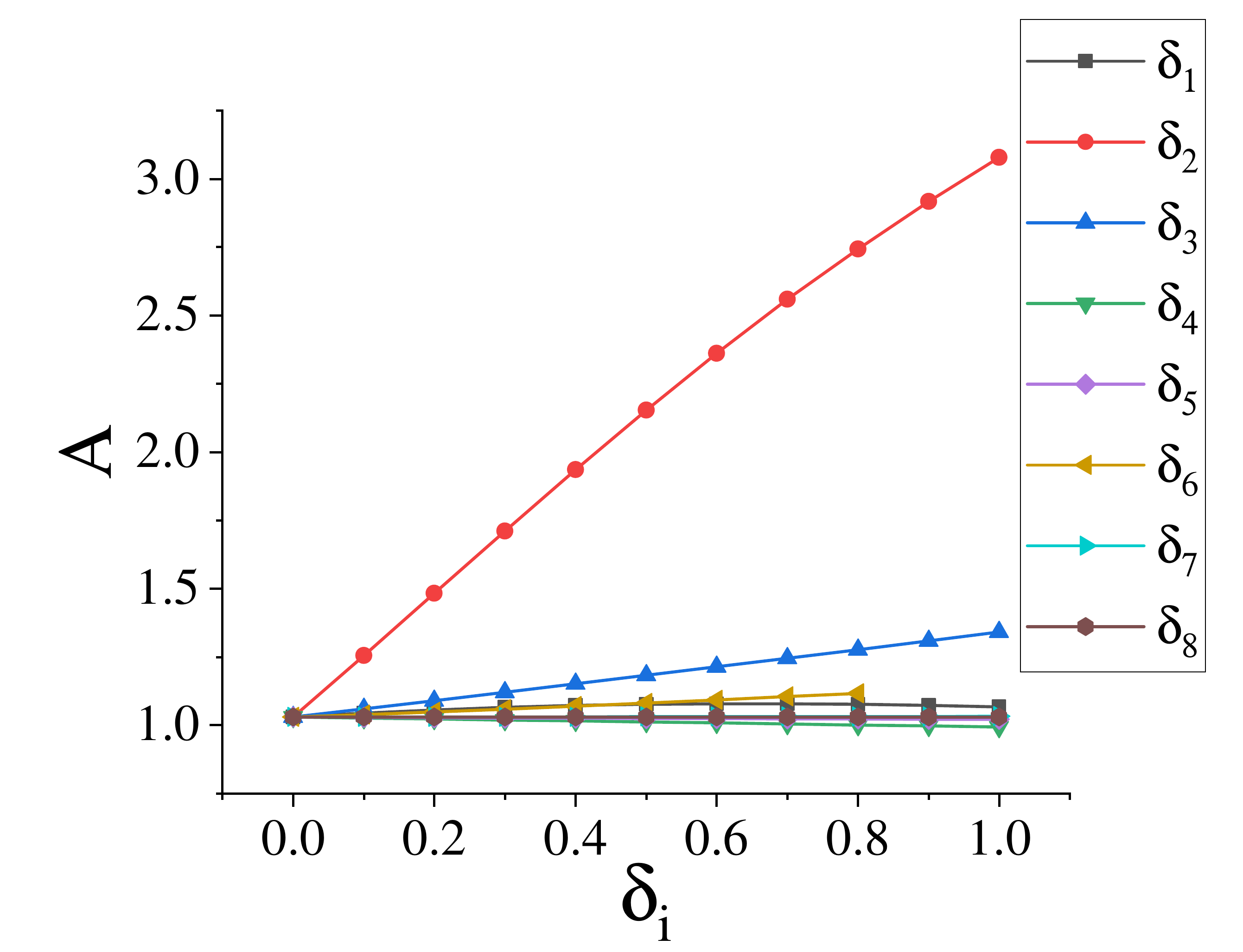}
}
\caption{Average radius $\langle R \rangle$ (top row) and asymmetry parameter $A$ (bottom row). The first column corresponds to the Kerr metric with the values of spin $0 \leq a \leq 1$, the second column corresponds to the KRZ metric with different values of $\delta_{i}$ and the spin $a=0.5$, when one of $\delta_{i}$ is confirmed, the other $\delta_{i}$ is set to be zero.   \label{RandA}}
\end{figure*}

In order to describe the dependence of the shadow's shape with different deformation parameters, we will use the coordinate independent formalism proposed in~\cite{Cele_coor_2}. The shape of the shadow is described by the horizontal displacement from the center of the image $D$, the average radius of the sphere $\langle R \rangle$, and the asymmetry parameter $A$. Since the KRZ spacetime is axially symmetric, the parameter $D$ is always identically equal to zero. In the papers Ref.~\cite{Ano_way_D_1,Ano_way_D_2} authors proposed different ways to describe the shape of the shadow. However, the results of the different approaches are similar to each other. The average radius $\langle R \rangle$ is the average distance of the boundary of the shadow from its center, which is defined by
\begin{eqnarray}
\langle R \rangle \equiv \frac{1}{2\pi} \int_{0}^{2\pi} R(\vartheta)d\vartheta,
\end{eqnarray}
where $R(\vartheta) \equiv \left[(\alpha-D)^2+\beta(\alpha)^2\right]^{1/2}$, $D=0$ and $ \vartheta \equiv \tan^{-1}[\beta(\alpha)/\alpha)]$. The asymmetry parameter $A$ is the distortion of the shadow from a circle and defined as
\begin{eqnarray}
A \equiv 2\left[\frac{1}{2\pi}\int_{0}^{2\pi} \left(R-\langle R \rangle \right)^2 d\vartheta \right]^{1/2}.
\end{eqnarray}

The shadow of the compact object in KRZ spacetime for the different values of metric parameters is shown in Figs.~\ref{shadow_Diff_metri}-\ref{delta8}. From the Figs.~\ref{shadow_Diff_metri}-\ref{delta8} one may come to the following conclusions:
\begin{itemize}
    \item  From Fig.~\ref{shadow_Diff_metri}~(a) one can see that with the increase of rotation parameter  $a$ one side of the shadow goes away from the center while other one comes closer. 
    
    \item From Fig.~\ref{shadow_Diff_metri}~(b) one can observe that the presence of the parameters  $\delta_{1}$ and $\delta_{2}$ force the shape of shadow to be more flatter.
    
    \item The Fig.~\ref{shadow_Diff_metri}~(c) shows the same effect similar to one caused by the rotation parameter $a$: the result is obvious because the effect of inclination angle between the observer lens axis and the normal of observer’s sky can be converted by projection of the spin $a$. 
    
    \item From Fig.~\ref{shadow_Diff_metri}~(d), one can see that the parameters  $\delta_{1}$ and $\delta_{2}$ don't affect inclination as much as they  spin parameter does.
    
    \item The Fig.~\ref{shadow_delta}~(a) shows the shadow is more small when $\delta_{1}$ increases. Since $\delta_{1}$ related to $g_{tt}$ one may conjecture as $g_{tt}$ increases the size of shadow decreases just like as the particle or photon moves around the black hole with different energy then captured by the black hole .
        
    \item The Fig.~\ref{shadow_delta}~(b) shows that with the increase of $\delta_{2}$ the shadow becomes more asymmetric and the center of the image will be shifted.  
    
    \item The Fig.~\ref{shadow_delta}~(c) shows the similar effect as one due to parameter $\delta_{2}$. Here the  interesting point is that with increase of the value of parameter $\delta_{3}$ the additional protrusions will appear in the image of the shadow. Since the effect of the $\delta_{2}$ and $\delta_{3}$ is similar to the one due to spin parameter $a$, they may be related to rotational deformations of the spacetime metric.
    
    \item The Fig.~\ref{shadow_delta}~(d)-(e) show that the size of shadow becomes bigger with the increase either $\delta_{4}$ or $\delta_{5}$. Due to this reason $\delta_{4}$ and $\delta_{5}$ may be related to deformations of $g_{rr}$. 
    
    \item The Fig.~\ref{shadow_delta}~(f) shows that 
    as the value of $\delta_{6}$ moves away from 0
     the shadow spreading inward/outward is more obvious, and the graphs become similar to the egrosphere's outer boundary. This may be due to fact that  the $\delta_{6}$ is related to deformations of the event horizon. 
    
\end{itemize}

Summarizing we may conclude that the parameters $\delta_{1}$ and $\delta_{4}$ can change the size of the shadow but with opposite effect, the parameters $\delta_{2}$ makes the shadow deviate from the center point, the parameters $\delta_{5}$ also can change the size of the shadow but not obvious and a little stretched, the parameters $\delta_{3}$ and $\delta_{6}$ can change the contour shape.

From Fig.~\ref{delta7} we can see when $\delta_{7}$ increases, the change of the shadow is small, the small picture in the upper right corner of the picture is the enlarged view of the graph. It shows the minor difference corresponding to the different values of parameter $\delta_{7}$. From Fig.~\ref{delta8} one can observe similar results corresponding to the parameter $\delta_{8}$. Since the effect of the $\delta_{7}$ and $\delta_{8}$ is very small, in some places (see, e.g. ~\cite{Ni_2016}) authors neglect the parameters  $\delta_{7}$ and $\delta_{8}$ and put the Eq.~(\ref{L}) to be 1.  Although the simplification has very weak impact on the shadow, it is mathematically misleading, because the metric used in the Ref.~\cite{Ni_2016} does not reduce to the Kerr metric when the $\delta_{i}=0$.

As for sensitivity, the shadow is sensitive to parameters $\delta_{1}$, $\delta_{2}$, $\delta_{4}$. The shadow is less sensitive to parameters
$\delta_{3}$, $\delta_{5}$, $\delta_{6}$. The shadow is almost not sensitive to parameters
$\delta_{7}$, $\delta_{8}$.

We also want to know how the KRZ metric can reproduce the exact metric, so we use the Kerr-Sen(KS) metric and Einstein Dilaton Gauss-Bonnet(EDGB) metric\cite{Younsi2016}. Fig.~\ref{shadow_ES_EDGB} shows the difference between the KS/EDGB metric and KRZ metric. Through this picture we can know KRZ metric has a good approximation to KS and EDGB metric.

Fig.~\ref{RandA} shows the dependence of $\langle R \rangle$ and $A$ as a function of the spin parameter (Kerr metric) and $\delta_{i}$ (KRZ metric) for the fixed values of inclination angle  $\theta_{0}=\pi/4$ and the spin parameter $a=0.5$. From Fig.~\ref{RandA}~(b) one can easily see that the $\delta_{1}$ and $\delta_{4}$ have a greater impact on $\langle R \rangle$, but with  different trend: when $\delta_{1}$ increases the value of $\langle R \rangle$ decreases and when $\delta_{4}$ increases the value of $\langle R \rangle$ increases. This is due to the fact that $\delta_{1}$ corresponds the deformation of $g_{tt}$, while $\delta_{4}$ corresponds the deformation of $g_{rr}$. On the other hand $\delta_{2}$ and $\delta_{3}$  have the same trend: when their values increase the $\langle R \rangle$ decreases. From Fig.~\ref{RandA}~(d), we may also see that the $\delta_{2}$ have a greater impact on $A$. Other $\delta_{i}$'s have small  influence on $A$. From Fig.~\ref{RandA}~(b)-(d), one may see that the trend of $\langle R \rangle$ and $A$ with $\delta_{4}$, $\delta_{5}$ always opposite to effect of $\delta_{2}$, $\delta_{3}$ and $\delta_{6}$. Finally, with the increase of $\delta_{7}$ and $\delta_{8}$, the value of $\langle R \rangle$ and $A$ change very slowly, which also shows that $\delta_{7}$ and $\delta_{8}$ are adjustment parameters and have weak effect on the KRZ metric.

\section{Conclusion\label{Summary}}
In the present work we have studied the photon motion in the equatorial plane around the generally axisymmetric black hole which is described by the parametrized KRZ metric. From the properties of the circular orbits of photon, we can quantify the frequency and decay rate of ringdown GW signals by the orbital frequency and Lyapunov exponent of the light-ring. At the same time, the shape of the shadow which be measured by distant observers is also gotten from the photons motion around the parametrized black hole. We have calculated the frequency and LE of the unstable circular orbits of photon in order to get the information on QNMs. We have also obtained the frequency and decay rate of ringdown which can be used to construct a waveform model for the KRZ black hole merger.

In the special case when the photon orbits in the equatorial plane, we have found that only two primary parameters $\delta_{1}$ and $\delta_{2}$ of the KRZ metric affect on photon trajectory. It has been shown that with the increase of the spin parameter $a$, $\delta_{1}$ and $\delta_{2}$ the radius of photon circular orbits decreases. However, the change of the radius of photon circular orbits are more sensitive to the change of spin parameter $a$, especially when the spin parameter exceeds the value $0.4$. It has been found out that effects of spin parameter $a$ and $\delta_{2}$ on the frequency of photon orbits is the same order while the effect of $\delta_{2}$ is weaker. With these fitted formula, one can in principle get the frequency of ringdown waveforms from a perturbed black hole described by the KRZ metric. 

The decay rate of QNM amplitude can be obtained from the LE characterizing the rate of divergence of nearby null geodesics. In this work we hve shown that when $\delta_{1}$ increases from 0 to 1, LE will have two maximum values and there is no obvious downward trend in the overall trend of the dependence. When $\delta_{2}$ increases from 0 to 1 LE will have one maximum value and the overall trend of the graph is down. This is related to the initial conditions of the equation of motion. Though the angular momentum $l$ increases with the decrease of $\delta_{1}$ and $\delta_{2}$. We have shown that the decay of ringdown is very sensitive to the KRZ parameters. 

We have also studied the shadow of the black hole described by the parametrized KRZ spacetime. It has been shown that  the parameters $\delta_{1}$ and $\delta_{4}$ can change the size of the shadow but with opposite effect. The parameter $\delta_{2}$ makes the shadow deviate from the center, the parameter $\delta_{5}$ also can change the size of the shadow but effect is relatively weak. The parameters $\delta_{3}$ and $\delta_{6}$ can change the contour shape. The shape of the shadow has different sensitivity to different parameters: the shadow is more sensitive to parameters $\delta_{1}$, $\delta_{2}$, $\delta_{4}$ and less sensitive to parameters
$\delta_{3}$, $\delta_{5}$, $\delta_{6}$. The effects of the parameters $\delta_{7}$ and $\delta_{8}$ to the shape of shadow is very weak and almost negligible (compare with~\cite{Ni_2016} where authors have neglected the effects of $\delta_{7}$ and $\delta_{8}$ on the iron line in the X-ray spectrum of black holes).

One of the main results of this paper is the analysis of the dependence of the average radius of the shadow and asymmetry (distortion) parameter from the spin parameter and KRZ parameters.  Among the effects of other parameters the effect of parameter $\delta_1$ is dominant in changing the average radius of the shadow. On the other hand the main contribution to the change of the asymmetry parameters comes due to the presence of the parameter $\delta_2$. In principle, one may see that the two observable quantities (radius of shadow and asymmetry parameter) could provide the rough estimation of the parameters $\delta_1$ and $\delta_2$. Further analysis of the KRZ prameters and comparison with particular black hole solutions may provide a useful tool to probe the gravity models.   

The ringdown and shadow reflect the dynamical process and geometric properties in the strong field of the black hole, respectively. The former can be observed by the ground and space-borne GW detectors and the later can be observed by the EHT. From the GWs and image of EHT, one can reveal the nature of the black holes and test if they are described by the Kerr spacetime which is an exact solution of Einstein field equation in general relativity and assumed to describe the astrophysical black holes. Fortunately, both of these two phenomenon are related to the photon orbits around the black hole at the light-ring. In the present work, by calculating the photon's motion at the light-ring, we qualify the QNMs and shadows of generally axisymmetric black holes under general parametrized metrics. Perturbing the supermassive black hole and radiating the ringdown signals can be expected in our Galaxy~\cite{han2020mnras}, and the imaging of this nearest supermassive black hole is a target of the EHT project. Our results may play a role to construct a joint constraint of dynamical and stationary spacetime of black hole with both LISA and EHT observations. 

\begin{acknowledgements}
This work is supported by NSFC No. 11773059, AA is supported through the PIFI fund of Chinese Academy of Sciences.  
\end{acknowledgements}

\appendix

\subsection*{Appendix subsection}

\section{}
The function describing the radius of the photon orbits in equatorial plane and under parametrized KRZ metric has the following form
\begin{align*}
F_{r}(\delta_{1},\delta_{2},\tilde{a})=&k+{{F}_{1}}(\tilde{a})+{{F}_{2}}({{\delta }_{1}})+{{F}_{3}}({{\delta }_{2}})+{{F}_{4}}({{\delta }_{1}},\tilde{a})
\\&+{{F}_{5}}({{\delta }_{2}},\tilde{a})+{{F}_{6}}({{\delta }_{1}},{{\delta }_{2}})+{{F}_{7}}({{\delta }_{1}},{{\delta }_{2}},\tilde{a})  
\end{align*}
\begin{align*}
K = 3.00209
\end{align*}
\begin{align*}
{{F}_{1}}(\tilde{a})=-1.28767\tilde{a}+0.34098{{\tilde{a}}^{2}}-0.76082{{\tilde{a}}^{3}}
\end{align*}
\begin{align*}
{{F}_{2}}({{\delta }_{1}})=-0.444022{{\delta }_{1}}+\text{0}\text{.13889}{{\delta }_{1}}^{2}
\end{align*}
\begin{align*}
{{F}_{3}}({{\delta }_{2}})=&\text{-1}\text{.40618}{{\delta }_{2}}\text{-9}\text{.94242}{{\delta }_{2}}^{2}\text{+37}\text{.6563}{{\delta }_{2}}^{3}\text{-53}\text{.23631}{{\delta }_{2}}^{4}\\&\text{+34}\text{.56977}{{\delta }_{2}}^{5}\text{-8}\text{.59813}{{\delta }_{2}}^{6}
\end{align*}
\begin{align*}
{{F}_{4}}\left( {{\delta }_{1}},\tilde{a} \right)=&0.0612867{{\delta }_{1}}\tilde{a}+1.14277{{\delta }_{1}}{{\tilde{a}}^{2}}-0.618507{{\delta }_{1}}{{\tilde{a}}}^{3}\\
&+0.6384\delta _{1}^{2}\tilde{a}-1.49899\delta _{1}^{2}{{\tilde{a}}^{2}}+0.59456\delta _{1}^{2}{{\tilde{a}}^{3}}\\
&-0.339307\delta _{1}^{3}\tilde{a}+0.573049\delta _{1}^{3}{{\tilde{a}}^{2}}-0.174293\delta _{1}^{3}{{\tilde{a}}^{3}}
\end{align*}
\begin{align*}
{{F}_{5}}({{\delta }_{2}},\tilde{a})=&\text{-1}\text{.45769}{{\delta }_{2}}\tilde{a}\text{+9.1916}{{\delta }_{2}}{{\tilde{a}}^{2}}\text{-5}\text{.08388}{{\delta }_{2}}{{\tilde{a}}^{3}} \\
&\text{+18}\text{.3726}{{\delta }_{2}}^{2}\tilde{a}\text{-40}\text{.509}{{\delta }_{2}}^{2}{{\tilde{a}}^{2}}\text{+20}\text{.9555}{{\delta }_{2}}^{2}{{\tilde{a}}^{3}} \\ 
&\text{-29}\text{.5518}{{\delta }_{2}}^{3}\tilde{a}\text{+55}\text{.2847}{{\delta }_{2}}^{3}{{\tilde{a}}^{2}}\text{-27}\text{.9031}{{\delta }_{2}}^{3}{{\tilde{a}}^{3}}\\
&\text{+13}\text{.799}{{\delta }_{2}}^{4}\tilde{a}\text{-24}\text{.2201}{{\delta }_{2}}^{4}{{\tilde{a}}^{2}}\text{+12}\text{.0752}{{\delta }_{2}}^{4}{{\tilde{a}}^{3}}
\end{align*}
\begin{align*}
{{F}_{6}}({{\delta }_{1}},{{\delta }_{2}})=&\text{-1}\text{.40231}{{\delta }_{1}}{{\delta }_{2}}\text{+21}\text{.6537}{{\delta }_{1}}{{\delta }_{2}}^{2}\text{-64}\text{.9458}{{\delta }_{1}}{{\delta }_{2}}^{3}\\&\text{+85}\text{.1184}{{\delta }_{1}}{{\delta }_{2}}^{4}\text{-52}\text{.1275}{{\delta }_{1}}{{\delta }_{2}}^{5}\text{+12}\text{.139}{{\delta }_{1}}{{\delta }_{2}}^{6}\\
& \text{+1}\text{.25238}{{\delta }_{1}}^{2}{{\delta }_{2}}\text{-11}\text{.7479}{{\delta }_{1}}^{2}{{\delta }_{2}}^{2}\text{+32}\text{.1809}{{\delta }_{1}}^{2}{{\delta }_{2}}^{3}\\&\text{-39}\text{.8198}{{\delta }_{1}}^{2}{{\delta }_{2}}^{4}\text{+22}\text{.9467}{{\delta }_{1}}^{2}{{\delta }_{2}}^{5}\text{-4}\text{.93156}{{\delta }_{1}}^{2}{{\delta }_{2}}^{6} \\ 
\end{align*}
\begin{align*}
{{F}_{7}}({{\delta }_{1}},{{\delta }_{2}},\tilde{a})=&\text{-4}\text{.30625}{{\delta }_{1}}{{\delta }_{2}}\tilde{a}\text{+27}\text{.7323}{{\delta }_{1}}^{2}{{\delta }_{2}}\tilde{a}\text{-20}\text{.6891}{{\delta }_{1}}^{3}{{\delta }_{2}}\tilde{a}\\
&\text{+49}\text{.0293}{{\delta }_{1}}{{\delta }_{2}}^{2}\tilde{a}\text{-212}\text{.63}{{\delta }_{1}}^{2}{{\delta }_{2}}^{2}\tilde{a}\text{+147}\text{.897}{{\delta }_{1}}^{3}{{\delta }_{2}}^{2}\tilde{a}\\
&\text{-103}\text{.967}{{\delta }_{1}}{{\delta }_{2}}^{3}\tilde{a}\text{+396}\text{.656}{{\delta }_{1}}^{2}{{\delta }_{2}}^{3}\tilde{a}\text{-269}\text{.764}{{\delta }_{1}}^{3}{{\delta }_{2}}^{3}\tilde{a}\\
&\text{+59}\text{.1911}{{\delta }_{1}}{{\delta }_{2}}^{4}\tilde{a}\text{-212}\text{.401}{{\delta }_{1}}^{2}{{\delta }_{2}}^{4}\tilde{a}\text{+142}\text{.897}{{\delta }_{1}}^{3}{{\delta }_{2}}^{4}\tilde{a}\\
&\text{-29}\text{.2685}{{\delta }_{1}}{{\delta }_{2}}{{\tilde{a}}^{2}}\text{+43}\text{.0102}{{\delta }_{1}}^{2}{{\delta }_{2}}{{\tilde{a}}^{2}}\text{-22}\text{.4658}{{\delta }_{1}}^{3}{{\delta }_{2}}{{\tilde{a}}^{2}}\\
&\text{+83}\text{.9201}{{\delta }_{1}}{{\delta }_{2}}^{2}{{\tilde{a}}^{2}}\text{-93}\text{.1165}{{\delta }_{1}}^{2}{{\delta }_{2}}^{2}{{\tilde{a}}^{2}}\text{+43}\text{.369}{{\delta }_{1}}^{3}{{\delta }_{2}}^{2}{{\tilde{a}}^{2}}\\
&\text{-75}\text{.7107}{{\delta }_{1}}{{\delta }_{2}}^{3}{{\tilde{a}}^{2}}\text{+31}\text{.359}{{\delta }_{1}}^{2}{{\delta }_{2}}^{3}{{\tilde{a}}^{2}}\text{-0}\\&\text{.681802}{{\delta }_{1}}^{3}{{\delta }_{2}}^{3}{{\tilde{a}}^{2}}\\
&\text{+19}\text{.9327}{{\delta }_{1}}{{\delta }_{2}}^{4}{{\tilde{a}}^{2}}\text{+20}\text{.2448}{{\delta }_{1}}^{2}{{\delta }_{2}}^{4}{{\tilde{a}}^{2}}\text{-20}\text{.796}{{\delta }_{1}}^{3}{{\delta }_{2}}^{4}{{\tilde{a}}^{2}}\\
&\text{+25}\text{.6721}{{\delta }_{1}}{{\delta }_{2}}{{\tilde{a}}^{3}}\text{-52}\text{.2594}{{\delta }_{1}}^{2}{{\delta }_{2}}{{\tilde{a}}^{3}}\text{+31}\text{.4421}{{\delta }_{1}}^{3}{{\delta }_{2}}{{\tilde{a}}^{3}}\\
&\text{-97}\text{.0986}{{\delta }_{1}}{{\delta }_{2}}^{2}{{\tilde{a}}^{3}}\text{+204}\text{.307}{{\delta }_{1}}^{2}{{\delta }_{2}}^{2}{{\tilde{a}}^{3}}\text{-125}\text{.38}{{\delta }_{1}}^{3}{{\delta }_{2}}^{2}{{\tilde{a}}^{3}}\\
&\text{+122}\text{.861}{{\delta }_{1}}{{\delta }_{2}}^{3}{{\tilde{a}}^{3}}\text{-260}\text{.441}{{\delta }_{1}}^{2}{{\delta }_{2}}^{3}{{\tilde{a}}^{3}}\text{+161}\text{.188}{{\delta }_{1}}^{3}{{\delta }_{2}}^{3}{{\tilde{a}}^{3}}\\
&\text{-50}\text{.8252}{{\delta }_{1}}{{\delta }_{2}}^{4}{{\tilde{a}}^{3}}\text{+107}\text{.8}{{\delta }_{1}}^{2}{{\delta }_{2}}^{4}{{\tilde{a}}^{3}}\text{-67}\text{.0748}{{\delta }_{1}}^{3}{{\delta }_{2}}^{4}{{\tilde{a}}^{3}}\\
\end{align*}\label{AppA}

\section{}
The function describing the frequency of the photon orbits in equatorial plane under parametrized KRZ metric has the form
\begin{align*}
F_{\omega}(\delta_{1},\delta_{2},\tilde{a})=&k+{{F}_{1}}(\tilde{a})+{{F}_{2}}({{\delta }_{1}})+{{F}_{3}}({{\delta }_{2}})+{{F}_{4}}({{\delta }_{1}},\tilde{a})
\\&+{{F}_{5}}({{\delta }_{2}},\tilde{a})+{{F}_{6}}({{\delta }_{1}},{{\delta }_{2}})+{{F}_{7}}({{\delta }_{1}},{{\delta }_{2}},\tilde{a})  
\end{align*}
\begin{align*}
K = 0.193
\end{align*}
\begin{align*}
{{F}_{1}}(\tilde{a})= 0.04952\tilde{a}+0.21633{{\tilde{a}}^{2}}-0.39405{{\tilde{a}}^{3}}+0.34292{{\tilde{a}}^{4}}
\end{align*}
\begin{align*}
{{F}_{2}}({{\delta }_{1}})=0.029013{{\delta }_{1}}+\text{0}\text{.0038367}{{\delta }_{1}}^{2}-\text{0}\text{.0014046}{{\delta }_{1}}^{3}
\end{align*}
\begin{align*}
{{F}_{3}}({{\delta }_{2}})=\text{0}\text{.04095}{{\delta }_{2}}\text{+0}\text{.47366}{{\delta }_{2}}^{2}\text{-0}\text{.17638}{{\delta }_{2}}^{3}
\end{align*}
\begin{align*}
{{F}_{4}}\left( {{\delta }_{1}},\tilde{a} \right)=&+0.0257427{{\delta }_{1}}\tilde{a}+0.0434687{{\delta }_{1}}{{\tilde{a}}^{2}}-0.0517708{{\delta }_{1}}{{\tilde{a}}}^{3}\\
&+0.00400781\delta _{1}^{2}\tilde{a}-0.0577969\delta _{1}^{2}{{\tilde{a}}^{2}}+0.0404687\delta _{1}^{2}{{\tilde{a}}^{3}}\\
&-0.00567736\delta _{1}^{3}\tilde{a}+0.0235175\delta _{1}^{3}{{\tilde{a}}^{2}}-0.0140544\delta _{1}^{3}{{\tilde{a}}^{3}}\\
\end{align*}
\begin{align*}
{{F}_{5}}({{\delta }_{2}},\tilde{a})=&\text{+0}\text{.348684}{{\delta }_{2}}\tilde{a}\text{+0.0161719}{{\delta }_{2}}{{\tilde{a}}^{2}}\text{-0}\text{.229635}{{\delta }_{2}}{{\tilde{a}}^{3}} \\
&\text{-0}\text{.413807}{{\delta }_{2}}^{2}\tilde{a}\text{-0}\text{.611906}{{\delta }_{2}}^{2}{{\tilde{a}}^{2}}\text{+0}\text{.508229}{{\delta }_{2}}^{2}{{\tilde{a}}^{3}} \\ 
&\text{-0}\text{.0418396}{{\delta }_{2}}^{3}\tilde{a}\text{+0}\text{.550562}{{\delta }_{2}}^{3}{{\tilde{a}}^{2}}\text{-0}\text{.326042}{{\delta }_{2}}^{3}{{\tilde{a}}^{3}} \\ 
\end{align*}
\begin{align*}
{{F}_{6}}({{\delta }_{1}},{{\delta }_{2}})=&\text{+0}\text{.7407}{{\delta }_{1}}{{\delta }_{2}}\text{-1}\text{.5587}{{\delta }_{1}}{{\delta }_{2}}^{2}\text{0}\text{.820459}{{\delta }_{1}}{{\delta }_{2}}^{3}\\
& \text{-1}\text{.6814}{{\delta }_{1}}^{2}{{\delta }_{2}}\text{+3}\text{.60301}{{\delta }_{1}}^{2}{{\delta }_{2}}^{2}\text{-1}\text{.93261}{{\delta }_{1}}^{2}{{\delta }_{2}}^{3}\\
& \text{+1}\text{.04502}{{\delta }_{1}}^{3}{{\delta }_{2}}\text{-2}\text{.2501}{{\delta }_{1}}^{3}{{\delta }_{2}}^{2}\text{+1}\text{.21019}{{\delta }_{1}}^{3}{{\delta }_{2}}^{3}\\
\end{align*}
\begin{align*}
{{F}_{7}}({{\delta }_{1}},{{\delta }_{2}},\tilde{a})
\\=&\text{+2}\text{.11264}{{\delta }_{1}}{{\delta }_{2}}\tilde{a}\text{-4}\text{.82564}{{\delta }_{1}}^{2}{{\delta }_{2}}\tilde{a}\text{+2}\text{.93892}{{\delta }_{1}}^{3}{{\delta }_{2}}\tilde{a}\\
&\text{+4}\text{.69601}{{\delta }_{1}}{{\delta }_{2}}^{2}\tilde{a}\text{-12}\text{.9595}{{\delta }_{1}}^{2}{{\delta }_{2}}^{2}\tilde{a}\text{+8}\text{.31046}{{\delta }_{1}}^{3}{{\delta }_{2}}^{2}\tilde{a}\\
&\text{-2}\text{.37167}{{\delta }_{1}}{{\delta }_{2}}^{3}\tilde{a}\text{+6}\text{.77483}{{\delta }_{1}}^{2}{{\delta }_{2}}^{3}\tilde{a}\text{-4}\text{.36464}{{\delta }_{1}}^{3}{{\delta }_{2}}^{3}\tilde{a}\\
&\text{-0}\text{.651905}{{\delta }_{1}}{{\delta }_{2}}{{\tilde{a}}^{2}}\text{-0}\text{.439267}{{\delta }_{1}}^{2}{{\delta }_{2}}{{\tilde{a}}^{2}}\text{+0}\text{.477762}{{\delta }_{1}}^{3}{{\delta }_{2}}{{\tilde{a}}^{2}}\\
&\text{-3}\text{.38313}{{\delta }_{1}}{{\delta }_{2}}^{2}{{\tilde{a}}^{2}}\text{+13}\text{.1134}{{\delta }_{1}}^{2}{{\delta }_{2}}^{2}{{\tilde{a}}^{2}}\text{-8}\text{.64233}{{\delta }_{1}}^{3}{{\delta }_{2}}^{2}{{\tilde{a}}^{2}}\\
&\text{+1}\text{.7363}{{\delta }_{1}}{{\delta }_{2}}^{3}{{\tilde{a}}^{2}}\text{-7}\text{.06456}{{\delta }_{1}}^{2}{{\delta }_{2}}^{3}{{\tilde{a}}^{2}}\text{+4}\text{.67207}{{\delta }_{1}}^{3}{{\delta }_{2}}^{3}{{\tilde{a}}^{2}}\\
&\text{+5}\text{.5584}{{\delta }_{1}}{{\delta }_{2}}{{\tilde{a}}^{3}}\text{-11}\text{.4069}{{\delta }_{1}}^{2}{{\delta }_{2}}{{\tilde{a}}^{3}}\text{+7}\text{.00828}{{\delta }_{1}}^{3}{{\delta }_{2}}{{\tilde{a}}^{3}}\\
&\text{-2}\text{.21971}{{\delta }_{1}}{{\delta }_{2}}^{2}{{\tilde{a}}^{3}}\text{+0}\text{.443027}{{\delta }_{1}}^{2}{{\delta }_{2}}^{2}{{\tilde{a}}^{3}}\text{-0}\text{.0293486}{{\delta }_{1}}^{3}{{\delta }_{2}}^{2}{{\tilde{a}}^{3}}\\
&\text{+0}\text{.846215}{{\delta }_{1}}{{\delta }_{2}}^{3}{{\tilde{a}}^{3}}\text{+0}\text{.77316}{{\delta }_{1}}^{2}{{\delta }_{2}}^{3}{{\tilde{a}}^{3}}\text{-0}\text{.622904}{{\delta }_{1}}^{3}{{\delta }_{2}}^{3}{{\tilde{a}}^{3}}\\
&\text{-3}\text{.60367}{{\delta }_{1}}{{\delta }_{2}}{{\tilde{a}}^{4}}\text{+7}\text{.94229}{{\delta }_{1}}^{2}{{\delta }_{2}}{{\tilde{a}}^{4}}\text{-4}\text{.94668}{{\delta }_{1}}^{3}{{\delta }_{2}}{{\tilde{a}}^{4}}\\
&\text{+2}\text{.08781}{{\delta }_{1}}{{\delta }_{2}}^{2}{{\tilde{a}}^{4}}\text{-3}\text{.02619}{{\delta }_{1}}^{2}{{\delta }_{2}}^{2}{{\tilde{a}}^{4}}\text{+1}\text{.85113}{{\delta }_{1}}^{3}{{\delta }_{2}}^{2}{{\tilde{a}}^{4}}\\
&\text{-0}\text{.823382}{{\delta }_{1}}{{\delta }_{2}}^{3}{{\tilde{a}}^{4}}\text{+0}\text{.819658}{{\delta }_{1}}^{2}{{\delta }_{2}}^{3}{{\tilde{a}}^{4}}\text{-0}\text{.488054}{{\delta }_{1}}^{3}{{\delta }_{2}}^{3}{{\tilde{a}}^{4}}\\
\end{align*}\label{AppB}

\section{}\label{appC}
When $\delta_{i} = 0$ the KRZ metric (\ref{metric}) reduces to the Kerr one. One can get the expression for $g_{rr}$ as 
$$g_{rr} =\Sigma \frac{B^{2}}{N^{2}}\ ,$$
where:
 \begin{align*}
\Sigma=&1+\frac{\tilde{a}^{2}}{r^{2}} \cos ^{2} \theta\\
B=&1+\delta_{4} r_{0}^{2} / \tilde{r}^{2}+\delta_{5} r_{0}^{2} \cos ^{2} \theta / \tilde{r}^{2}
\end{align*}
\begin{align}\label{N2_KRZR_AppC}
{N}^{2}=&\left( 1-{{r}_{0}}/\tilde{r} \right)\nonumber\\
&\left[ 1-{{\epsilon }_{0}}{{r}_{0}}/\tilde{r}+\left( {{k}_{00}}-{{\epsilon }_{0}} \right)r_{0}^{2}/{{{\tilde{r}}}^{2}}+{{\delta }_{1}}r_{0}^{3}/{{{\tilde{r}}}^{3}} \right]\nonumber\\
&+[ {{a}_{20}}r_{0}^{3}/{{{\tilde{r}}}^{3}}+{{a}_{21}}r_{0}^{4}/{{{\tilde{r}}}^{4}}+{{k}_{21}}r_{0}^{3}/{{{\tilde{r}}}^{3}}L]{{\cos }^{2}}\theta
\end{align}
when $\delta_{i} = 0$, then we get 
\begin{eqnarray}
N^{2} &=&  (a^2+\tilde{r}^2-2\tilde{r})/\tilde{r}^2\ ,\\ 
B&=&1\ ,\\
\Sigma(r, \theta)&=&1+\frac{\tilde{a}^{2}}{\tilde{r}^{2}} \cos ^{2} \theta \ ,
\end{eqnarray}  
and
\begin{equation}
 g_{rr} = \frac{\tilde{r}^{2}+\tilde{a}^2 \cos ^{2} \theta } {\tilde{r}^2-2\tilde{r}+\tilde{a}^2}\ ,
\end{equation}
which coincides the metric component $g_{rr}$ in Kerr metric with the unit mass $M =1$.

However in~\cite{Ni_2016} authors have written $N^{2}$ in the form:
\begin{align}\label{N2_KRZW_AppC}
N^{2}=&\left(1-\frac{r_{0}}{\tilde{r}}\right)\left[1-\frac{\epsilon_{0} r_{0}}{\tilde{r}}+\left(k_{00}-\epsilon_{0}\right) \frac{r_{0}^{2}}{\tilde{r}^{2}}+\frac{\delta_{1} r_{0}^{3}}{\tilde{r}^{3}}\right]\nonumber\\
&+\left[\left(k_{21}+a_{20}\right) \frac{r_{0}^{3}}{\tilde{r}^{3}}+\frac{a_{21} r_{0}^{4}}{\tilde{r}^{4}}\right] \cos ^{2} \theta
\end{align}

From (\ref{N2_KRZR_AppC}) and (\ref{N2_KRZW_AppC}), one can see the difference between the parameter $N^{2}$ is the function in front of the $\cos ^{2} \theta$. One can simplify the expression (\ref{N2_KRZW_AppC}) and after simple calculation one may find the function in front of the $\cos ^{2} \theta$ in the form $$[a^4/(\tilde{r}^3((1 - a^2)^(1/2) + 1)) - a^4/\tilde{r}^4],$$
so (\ref{N2_KRZW_AppC}) takes the form 
\begin{equation} \label{S}
N^{2}=(a^2 + \tilde{r}^2 - 2\tilde{r})/\tilde{r}^2+[a^4/(\tilde{r}^3((1 - a^2)^(1/2) + 1)) - a^4/\tilde{r}^4]\cos ^{2} \theta\ .
\end{equation}
When we put the function (\ref{S}) into the function $g_{rr}$, we find that the $g_{rr}$ can not reduce to the $g_{rr}$ for Kerr metric. However, since the parameters $\delta_{7}$ and $\delta_{8}$ have a small impact to the spacetime, one may neglect these parameter while using the KRZ paramertrization.

\bibliographystyle{apsrev4-1}  %% BibTeX style

\end{document}